\documentclass[screen, acmlarge]{acmart}

\AtBeginDocument{%
  \providecommand\BibTeX{{%
    \normalfont B\kern-0.5em{\scshape i\kern-0.25em b}\kern-0.8em\TeX}}}

\setcopyright{rightsretained}

\acmSubmissionID{7660}

\usepackage{subcaption}
\usepackage[]{algorithm2e}
\usepackage{multirow}

\usepackage{url}
\makeatletter
\g@addto@macro{\UrlBreaks}{\UrlOrds}
\makeatother

\usepackage{colortbl}

\usepackage{multirow} 
\usepackage{color} 
\usepackage{enumitem} 

\usepackage{xspace}
\usepackage{xcolor, colortbl}

\usepackage{arydshln}
\usepackage{pifont}

\usepackage[final]{pdfpages}
\usepackage[]{nth}

\newcommand{\m}{\textit{M=}}

\newcommand{\sd}{\textit{SD=}}

\newcommand{\N}{\textit{N=}}

\newcommand{\rg}[2]{\textit{range=[#1, #2]}}
\newcommand{\greencheck}{{\color{green}\ding{52}}}
\newcommand{\redxmark}{{\color{red}  \ding{55}}}

\newcommand{\tool}{\textsc{AutoTherm}\xspace}
\newcommand{\toolData}{\textsc{AutoTherm dataset}\xspace}
\newcommand{\toolDatas}{\textsc{AutoTherm datasets}\xspace}
\newcommand{\indoor}{\textsc{indoor}\xspace}
\newcommand{\indoorDataset}{\indoor dataset\xspace}
\newcommand{\vehicle}{\textsc{vehicle}\xspace}
\newcommand{\vehicleDataset}{\vehicle dataset\xspace}

\newcommand{\seven}{$\kappa_7$\xspace}
\newcommand{\three}{$\kappa_3$\xspace}
\newcommand{\two}{$\kappa_2$\xspace}

\newcommand{\minmax}[4]{[\textcolor{#1}{{#3}}, \textcolor{#2}{{#4}}]}
\newcommand{\meanstd}[2]{$\mathrel{{\mathrel{{{#1}}{\pm}}}{{#2}}}$}

\definecolor{bluegray}{HTML}{4e79a7}
\definecolor{darkorange}{HTML}{f28e2c}
\definecolor{darkpastelred}{HTML}{e15759}
\definecolor{etonblue}{HTML}{76b7b2}

\def\plaintitle{AutoTherm: A Dataset and Benchmark for Thermal Comfort Estimation Indoors and in Vehicles}

\def\plainkeywords{Machine Learning; vehicles; state recognition; dataset.}

\begin{document}

\title{\plaintitle}

\author{Mark Colley}
\authornote{Both authors contributed equally to this research.}
\email{mark.colley@uni-ulm.de}
\orcid{0000-0001-5207-5029}
\affiliation{%
  \institution{Institute of Media Informatics, Ulm University}
  \city{Ulm}
  \country{Germany}
}

\author{Sebastian Hartwig}
\authornotemark[1]
\email{sebastian.hartwig@uni-ulm.de}
\orcid{0000-0001-8642-2789}
\affiliation{%
  \institution{Institute of Media Informatics, Ulm University}
  \city{Ulm}
  \country{Germany}
}

\author{Albin Zeqiri}
\email{albin.zeqiri@uni-ulm.de}
\orcid{0000-0001-6516-3810}
\affiliation{%
  \institution{Institute of Media Informatics, Ulm University}
  \city{Ulm}
  \country{Germany}
}

\author{Timo Ropinski}
\email{timo.ropinsik@uni-ulm.de}
\orcid{0000-0002-7857-5512}
\affiliation{%
  \institution{Institute of Media Informatics, Ulm University}
  \city{Ulm}
  \country{Germany}
}

\author{Enrico Rukzio}
\email{enrico.rukzio@uni-ulm.de}
\orcid{0000-0002-4213-2226}
\affiliation{%
  \institution{Institute of Media Informatics, Ulm University}
  \city{Ulm}
  \country{Germany}
}
\renewcommand{\shortauthors}{Colley and Hartwig et al.}

\begin{abstract}
Thermal comfort inside buildings is a well-studied field where human judgment for thermal comfort is collected and may be used for automatic thermal comfort estimation. However, indoor scenarios are rather static in terms of thermal state changes and, thus, cannot be applied to dynamic conditions, e.g., inside a vehicle. In this work, we present our findings of a gap between building and in-vehicle scenarios regarding thermal comfort estimation. We provide evidence by comparing deep neural classifiers for thermal comfort estimation for indoor and in-vehicle conditions. Further, we introduce a temporal dataset for indoor predictions incorporating $31$ input signals and self-labeled user ratings by $18$ subjects in a self-built climatic chamber. For in-vehicle scenarios, we acquired a second dataset featuring human judgments from $20$ subjects in a BMW $3$ Series. Our experimental results indicate superior performance for estimations from time series data over single vector input. Leveraging modern machine learning architectures enables us to recognize human thermal comfort states and estimate future states automatically. We provide details on training a recurrent network-based classifier and perform an initial performance benchmark of the proposed dataset. Ultimately, we compare our collected dataset to publicly available thermal comfort datasets. 
\end{abstract}



\begin{teaserfigure}
  \includegraphics[width=\textwidth]{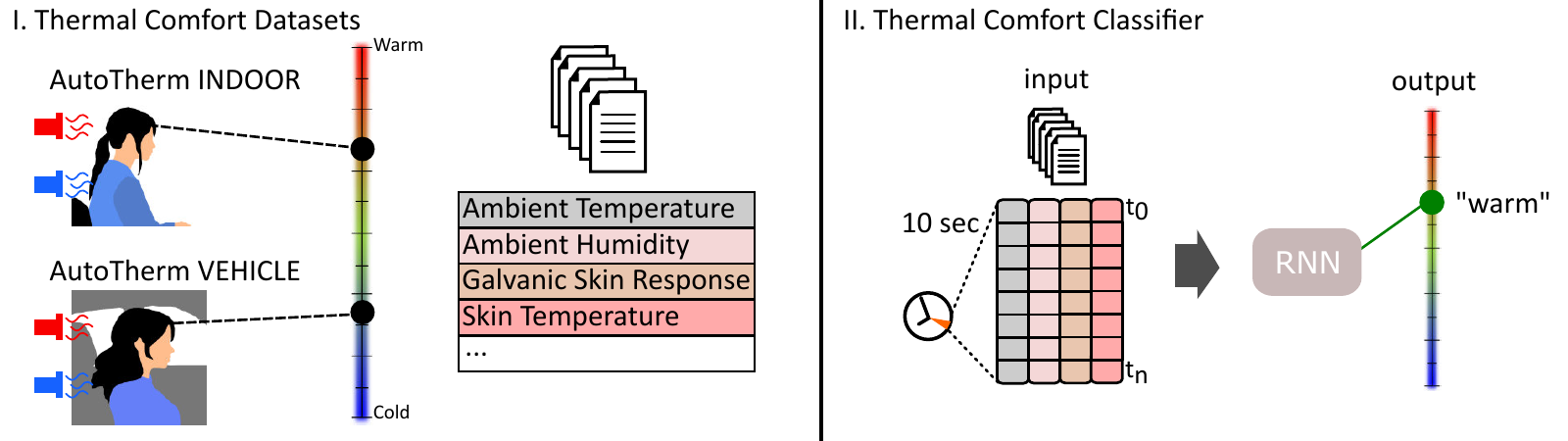}
  \caption{We propose the \toolDatas. (I.) We collected human-annotated sensor measurements during two thermal state studies (\indoor and \vehicle), enabling the training of neural thermal state classifiers (II.) to identify human thermal state changes from sequential sensor measurements.}
  \Description{This Figure shows the two stages of our dataset introduction. On the left side, a person is depicted with two fans blowing warm or cold air. This leads to a thermal comfort rating, indicated via a scale from warm to cold. The figure then shows the various measurements, such as ambient temperature. On the right side, the figure shows via a clock that 10 sec time steps were used to predict the thermal comfort.}
  \label{fig:teaser}
\end{teaserfigure}

\maketitle

\section{Introduction}
The inclusion of sensors such as cameras, radars, thermometers, or lidars in today's manually driven vehicles and, most likely, in future (automated) vehicles~\cite{sensory} allows for novel insights into users' states and intentions. The same can be said of modern smart home systems, which automate indoor climate conditions and optimize energy efficiency. To improve user experience, there is a need for accurate and reliable recognition, interpretation, and understanding of current user states~\cite{9597393}, as this ensures the execution of adjustments based on user preferences~\cite{stampf2022towards} and alleviates the user of burdensome tasks such as adjusting the temperature. Additionally, manually adjusting the temperature can lead to an overshoot~\cite{petre2019automatic}, requiring additional interactions. 
Using in-vehicle sensors and machine-learning methods, current vehicles already recognize some driver states, such as level of drowsiness~\cite{manstettenEvolutionDriverMonitoring2020} or fatigue~\cite{fatigueSurvey}. This is mainly a safety measure to avert potentially dangerous driving behavior. Yet, seeing as driving-related tasks will increasingly become irrelevant the higher the level of automation, recognition of other states such as emotional state~\cite{bethge} and intention~\cite{nilssonActionIntentionRecognition} become increasingly relevant. There are various methods to determine the passenger's state, including machine-learning-based methods. Depending on whether the state recognition task is formulated as a supervised or unsupervised learning problem, labeled data is required~\cite{sathyaComparisonSupervisedUnsupervised2013}. Learning from human judgments to replicate human behavior or to mirror human preferences is a common practice~\cite{wu2023human, xu2023imagereward, lee2023aligning,hartwig2022learning, hartwig2023clusternet}, also indoors~\cite{jayathissa2020humans}. However, there are only a few labeled and publicly available datasets for the automotive state recognition use case, such as \textit{drive\&act} by \citet{drive&act} or the \textit{VEmotion dataset} by \citet{bethge}. While there are existing datasets for indoor thermal comfort, they cannot be applied to the field of in-vehicle environments.


Already today, one relevant aspect for passengers inside vehicles is the perceived level of thermal comfort~\cite{obornePassengerComfortOverview1978, rommelfanger2020evaluation, 10.1145/3534617}. In building ergonomics, thermal comfort, and its influencing factors have been part of numerous research studies that resulted in different models for thermal comfort estimation, such as the commonly known predicted mean vote (PMV) index by \citet{fanger1970thermal}. However, while thermal comfort is dependent on the exposure history~\cite{VELT201742}, these works focus on a single data point while we also incorporate temporal data. Additionally, there are significant differences between indoor settings and vehicles and the required data.
First, indoors, and especially offices, occupants do not have direct control access to the temperature. Second, this temperature changes only slowly. Therefore, the available datasets do not include temporal data but only provide singular data points. Additionally, the data points are intended to assess current comfort without the explicit goal of adapting the temperature. The automotive use case, on the other hand, is defined by possibly fast-changing temperatures (e.g., due to opened windows also by other users), limited variety and action performed by the users, and the current expectation of having the personally optimal temperature due to relatively easy access.
Additionally, in the context of automotive state recognition systems, thermal comfort state estimation (i.e., \textit{warm, comfortable, cold, ...}) has thus far not been explored. Consequently, no machine-learning model or dataset exists for in-vehicle thermal comfort state estimation. The impact of inputs from in-vehicle sensory units on prediction performance is also unknown.
The automotive domain is particularly characterized by providing an environment known by manufacturers (regarding size, capabilities, limitations, and heating possibilities), providing (already today) a plethora of sensor data, and restricting the user's movements and actions. Therefore, a specialized dataset is required.

\textit{Contribution Statement:} In this work, we created a (1) \textbf{temporal}, labeled multi-modal dataset featuring \textbf{31} input signals (age, gender, weight, height, body fat, body core temperature, activity level, time since last meal, tiredness, clothing level, radiation temperature, emotion, RGB frame, ten body pose key points, heart rate, wrist temperature, galvanic skin response, ambient temperature, relative humidity) relevant for the task of thermal comfort state estimation, which has not yet been addressed in the context of automotive state recognition. (2) The implemented and employed logging can be used as a data-gathering blueprint for future state recognition research. (3) Thirdly, we present a machine-learning-based approach for both indoor and in-vehicle thermal comfort state recognition that takes advantage of different feature combinations to explore predictive performance and the impact of individual input modalities. (4) Fourthly, we report a feature combination study with different network architectures for thermal comfort state estimation and forecasting. (5) Finally, we evaluate our trained classifiers on existing thermal comfort datasets and report superior performance for models trained on our \toolData.

Within the remainder of this paper, we will first discuss the work related to our approach in Section~\ref{sec:related_work}. Second, we're providing details on our study setup, where we collect human judgments for thermal comfort in Section~\ref{sec:thermal_comfort_study}, before describing the data we collected for indoor scenarios, in Section~\ref{sec:indoor_dataset} and for in-vehicle scenarios in Section~\ref{sec:vehicle_dataset}. We then compare our dataset to existing thermal comfort datasets, in Section~\ref{sec:datasets_comparison} followed by Section~\ref{sec:experiments} where we conduct and present results of several thermal comfort estimation experiments. Finally, we discuss our findings and address limitations of our approach in Section~\ref{sec:discussion}, share our dataset and code in Section~\ref{sec:availability} and conclude in Section~\ref{sec:conclusion}.

\section{Related Work}\label{sec:related_work}
This work builds on previous work on indoor thermal comfort, factors influencing thermal comfort, and state recognition in general.

\subsection{Thermal Comfort}
Various factors influence one's perceived comfort level in an indoor environment, such as visual, acoustic, and environmental conditions~\cite{FRONTCZAK2011922}. Thermal comfort describes the level of satisfaction with one's surroundings based on thermal influences~\cite{ramspeckASHRAESTANDARDSCOMMITTEE} and has been extensively researched in the field of building ergonomics. The PMV index \cite{fanger1970thermal} referenced in the ISO 7730:2006-05 and American Society of Heating, Refrigerating and Air-Conditioning Engineers (ASHRAE) 55-2020 standards~\cite{ISO7730, ramspeckASHRAESTANDARDSCOMMITTEE} estimates the perceived level of thermal comfort for a large group of people on a thermal sensation scale with the seven items \textit{Cold, Cool, Slightly Cool, Comfortable, Slightly Warm, Warm, and Hot} and is based on empirical thermal comfort studies, from which an equation for thermal comfort calculation based on six main influencing factors was derived: Metabolic Rate, Clothing Insulation, Mean Radiation Temperature, Ambient Temperature, Relative Humidity, and Air Velocity.

\subsection{Influences on Thermal Comfort}\label{sec:rw2}
Research on building ergonomics indicates that environmental~\cite{parsonsHumanThermalEnvironments2002} or physiological (human thermoregulation~\cite{mechanisms}) factors and their interplay should be considered. Thermal sensation is mostly felt due to thermoreceptors on one's skin and muscles~\cite{terjungComprehensivePhysiology2011}. Accordingly, skin temperature has been used as an indicator of thermal perception changes (e.g., \cite{ramanathanNewWeightingSystem1964, simEstimationThermalSensation2016}). When used in conjunction with body core temperature, \citet{InfluenceOfBodyTemp1999} found that skin and body core temperature contribute similarly to thermal comfort. \citet{simEstimationThermalSensation2016} demonstrate the estimation of thermal comfort based on measuring different sites around the wrist and fingertips, and \citet{ramanathanNewWeightingSystem1964} proposed an approach for the estimation of the mean skin temperature, computed by averaging the skin temperature of different locations across the body. Another work proposed an estimation of thermal comfort from multiple physiological input streams, such as heart rate, skin temperature, or electrodermal activity~\cite{yoshikawaCombiningThermalCamera2019}. Additionally, it was established that there is a range in which occupants feel thermally comfortable (thermal comfort zone)~\cite{CIUHA2016123, ciuhaThermalComfortZone2017}, rather than a single temperature. This zone is influenced by the dynamics of temperature change and the direction of the change~\cite{ciuhaEffectThermalTransience2019}. Apart from estimation based on physiological input data, it was reported that there is an influence of gender~\cite{CHAUDHURI2018391, karjalainenThermalComfortGender2012}, age~\cite{DELFERRARO2015177, GUERGOVA201180}, and emotion~\cite{WANG2020109789} on thermal comfort perception. For instance, female occupants seem more sensitive to thermal changes that deviate from their optimal state and thus feel too cold or too hot more frequently~\cite{SAMI20071594}. In the elderly, deterioration of skin receptors is assumed to cause reduced thermal perception ability, especially in the limbs~\cite{GUERGOVA201180}. As for emotion, \citet{WANG2020109789} concluded that negative emotions have an unfavorable effect on thermal comfort. However, overall, emotions only affect thermal comfort perception during light activities such as sitting or standing.

\subsection{Thermal Comfort Label Scale}\label{ssec:thermal_comfort_scales}
Fanger's PMV index~\cite{fanger1970thermal} is calculated in the interval [-3, 3]. Accordingly, the seven different thermal comfort states \textit{Cold, Cool, Slightly Cool, Comfortable, Slightly Warm, Warm, and Hot} represent ranges in the defined interval rather than integers. For instance, according to the ASHRAE standard~\cite{ramspeckASHRAESTANDARDSCOMMITTEE}, the state \textit{Comfortable} is established in the interval [-0.5, 0.5]. This more granular approach allows for reducing the initial seven-point scale to a three-point scale, where the new reduced states can be denoted as \textit{Too Cold, Comfortable, Too Warm}. Due to its simplicity and standardized theoretical basis, the seven-point thermal sensation scale was adopted as the label set.

\subsection{State Recognition Systems}\label{sec:rw1}
Machine-learning-based state recognition was used for cognitive load detection~\cite{cogload, yincogload}, driver stress detection~\cite{muhammad2023contactless}, situation awareness prediction~\cite{kim2023physiological}, affective computing~\cite{levasa21, emotionmeter}, vehicle assistance systems~\cite{manstettenEvolutionDriverMonitoring2020, lane, taoAssessmentDriversComprehensive2020}, and even pain recognition~\cite{walterDataFusionAutomated2015}. They mainly differ in terms of chosen input spaces, modalities, and employed learning methods (supervised or unsupervised learning). 
Even though systems trained on data from a single input space or single modality can perform satisfyingly, a closer resemblance to human perception can be achieved by incorporating additional modalities~\cite{multimodalMachineLearning}. Therefore, multi-modal input data for the training of state recognition systems is beneficial.\\
Early works in multi-modal in-vehicle state recognition incorporated various vehicle parameters (pedal position, steering), environmental information (local and global vehicle position), and driving performance attributes (speed) as input signals~\cite{berndtContinuousDriverIntention2008, heDrivingIntentionRecognition2012} to improve intention recognition in safety systems. Likewise, human action recognition experienced advances by incorporating multi-modal input signals together with feature fusion and co-learning methods~\cite{HAR}. \citet{zhangDriverBehaviorRecognition2020} demonstrated in-vehicle action recognition with their proposed interwoven CNN approach and a self-recorded dataset. Additionally, within the field of affective computing, the recognition of emotional states was explored. This was done as multi-modal interfaces, not just in the automotive context, could benefit from the ability to recognize and interpret one's emotions. With VEmotion, \citet{bethge} proposed a novel way of estimating the emotional state of drivers in real-time using driving context information such as weather, traffic, road, and car trajectory data. They demonstrated that states such as emotions can be predicted by using mainly contextual information, which is more readily available in vehicles. Generally, incorporation of insights from areas, such as emotion recognition~\cite{levasa21, emotionmeter}, cognitive load estimation~\cite{cogload} or next interaction method prediction~\cite{wolf21} can contribute to creating a more holistic understanding of users' needs in automotive state recognition systems.\\
Nevertheless, only a few publicly available datasets can be used for further state recognition research. Therefore, developing new approaches for recognizing certain states almost always entails the acquisition of a new dataset, thereby significantly slowing down development speed while increasing task complexity. Additionally, methods for in-vehicle thermal comfort estimation from multi-modal data have not been explored, although thermal sensation was, in other settings, reported to be one of the primary influencing factors in overall comfort perception~\cite{FRONTCZAK2011922}.

\subsection{Thermal Comfort Estimation}\label{sec:rw3}
Automated recognition of thermal comfort levels has been a research focus in energy and building ergonomics. For example, energy efficiency in office buildings or other occupant spaces could be improved through adaptation to occupants' current needs~\cite{veselyPersonalizedConditioningIts2014}. Furthermore, adaptive models for personalized thermal comfort were also correlated with an increase in occupant productivity \cite{buenoEvaluatingConnectionThermal2021} as environmental factors are better tailored to occupants' needs. 
Previous data-driven approaches for thermal comfort estimation showed differences in the used datasets (self-recorded or publicly available). With a self-recorded dataset featuring data from 12 participants, \citet{zhangFrownBasedThermalComfort2021} developed a building context machine-learning model that estimates thermal sensation with an accuracy of 95.4\% by identifying frowning facial expressions in recorded RGB data and relating the occurrences to the currently perceived thermal sensation. \citet{maoThermalComfortEstimation2021}concluded that thermal comfort could be predicted using heart rate and left-arm wrist skin temperature measures by using a self-recorded dataset to train different machine-learning models. As part of the advances in thermal comfort research, gathering sufficiently large datasets for data-driven thermal comfort estimation has become a research focus, resulting in the acquisition of the ASHRAE RP-884~\cite{ashrae1} and, more recently, the ASHRAE II \cite{ashrae2} datasets. Both datasets accumulate environmental, personal attributes, and thermal indices, one of which is the PMV index. Moreover, both ASHRAE datasets are publicly available. Using the ASHRAE RP-884~\cite{ashrae1}, Scales~\cite{scales_project}, and US Office Buildings dataset~\cite{us_office} (all public), \citet{somuHybridDeepTransfer2021} built a machine-learning model that employs transfer learning strategies and achieved a prediction accuracy of 55\%. The Scales Project dataset is a cross-national dataset (30 countries) that explores the occupants' understanding of common thermal sensation scales, such as the previously described seven-point scale. It includes thermal comfort labels based on different rating scales for thermal conditions. Additionally, personal, indoor, and outdoor environmental factors were gathered using a questionnaire. The US Office Buildings dataset is aimed at office spaces in the US and was recorded to explore human-building interactions driven by factors such as comfort and behavioral changes over time~\cite{us_office}. It includes data on personal attributes, indoor/outdoor variables, and labels gathered with various thermal comfort rating scales. 

\citet{10.1145/3360322.3360858} presented OccuTherm, a system to estimate thermal comfort using the body shape. They conducted a sensing study in which biometrics, physical measurements (height, shoulder circumference), and subjective comfort responses were recorded. They find that an adapted personalized comfort model can improve model performance to $60\%$ accuracy.

\citet{10.1145/3485730.3493693} collected a longitudinal dataset of $17$ participants over four weeks across $17$ indoor and outdoor spaces. The dataset includes physical characteristics, background information, and personality surveys, which were assessed once. During the four-week trial, thermal preference, clothing level, metabolic rate, perceived air velocity, and location were assessed. The dataset contribution contains $1.400$ unique responses across $17$ indoor and outdoor spaces.

\citet{QUINTANA2023109685} also introduced a cohort comfort model to reduce the necessity for personalized data. While not achieving the accuracy of personalized models, this approach opens novel ways to estimate thermal comfort without additional data. 



An overview of publicly available datasets is given in \autoref{tab:comparison_1}. Given the low number of public datasets (six in total), most thermal comfort research is still performed on self-recorded datasets that vary in terms of included measures and employed sensory devices. Additionally, no thermal comfort dataset for the automotive use case currently exists.

\section{Method - Thermal Comfort Study}\label{sec:thermal_comfort_study}
This work investigates thermal comfort estimation from temporal data in two scenarios, indoors and in vehicles. To assess thermal comfort for such scenarios, we conducted two studies enabling us to collect human judgments for varying thermal conditions in each scenario. For the indoor scenario, we set up a climate chamber, enabling us to control thermal conditions carefully. However, the study inside a vehicle is conducted with less control, utilizing a BMW 3 Series to capture the thermal properties of real in-vehicle conditions. In this section, we provide details on how we collected human feedback during the study using our developed application. For each subject, we gathered demographics using a self-report integrated into our application.

Self-reports and labels were gathered using a GUI that incorporates various dialogues. A keyboard was used for interaction because it requires very little space and does not require elaborate hand movements. 
During the trials, a feeling of boredom quickly manifested due to the repetitive task of simply labeling one's thermal comfort level. 
Consequently, for the indoor study, the waiting dialogue between labeling prompts was extended to display a slide show of fractal images changing every $2$ seconds. A dialog asking users to provide a rating on the $7$-point thermal comfort scale was displayed with an interval of $20$ seconds.
While we could not provide an acclimatization period between the heating and cooling period, we let participants first fill out a demographic questionnaire and explained the scenario. This took approximately $20$ min, providing sufficient time for initial acclimatization of the climate chamber. Nonetheless, this remains a limitation of our indoor dataset.

The experimental procedure followed the guidelines of the ethics committee of our university and adhered to regulations regarding the handling of sensitive and private data, anonymization, compensation, and risk aversion. Compliant with our university's local regulations, no additional formal ethics approval was required.

\section{Thermal Comfort Indoor Dataset}\label{sec:indoor_dataset}
For the exploration of temporal data in conjunction with thermal comfort estimation and to provide a scientific comparison of human thermal comfort between indoor scenarios and in-vehicle scenarios, within this work, we collect human judgments for both scenarios. In this section, we provide a detailed description of our \textbf{\indoorDataset} acquisition. The \indoorDataset represents scenarios where participants are sitting, which constitutes almost 70\% of a typical work day in the office and 60\% at home, thereby showing the relevance and appropriateness of the scenario~\cite{clemes2014office}. \\

Gathering ratings for thermal comfort states can be done using the thermal sensation scale referenced in the ASHRAE standard~\cite{ramspeckASHRAESTANDARDSCOMMITTEE}, which is defined in the same range as the PMV output scale proposed by \citet{fanger1970thermal}. The thermal sensation scale comprises seven-string encodings \textit{Cold, Cool, Slightly Cool, Comfortable, Slightly Warm, Warm, Hot} but can be represented numerically as $[-3, -2, -1, 0, +1, +2, +3]$. Additionally, a significant attribute of the PMV index is its scope, as it was designed to predict thermal comfort in steady-state environments~\cite{ISO7730, fanger1970thermal}, 
yet, temperatures inside buildings can vary depending on the context, seasonal climate, geographical climate, architectural properties, heating/cooling system, etc. To simulate the temperature of buildings, these factors need to be considered for data acquisition. In particular, the rate of change is important as thermal comfort zones in cooling and heating phases were found to differ, especially with smaller temperature step changes~\cite{ciuhaEffectThermalTransience2019}. A slow rate of temperature change in previous work was defined as $0.5$ °C/min, while a fast change rate was defined as $1.0$ °C/min~\cite{ciuhaEffectThermalTransience2019}. Temperature ranges used in thermal comfort experiments included ranges such as [$18$°,$35$°C]~\cite{simEstimationThermalSensation2016} or even larger ranges with a minimum of $15$°C and a maximum of $40$°C~\cite{ciuhaEffectThermalTransience2019}. Consequently, the temperature ranges and step sizes should be selected so that the conditions for all possible thermal comfort states are met at least once within the heating and cooling phases. 

\begin{figure*}[ht!]
    \centering
      \begin{subfigure}[c]{0.48\textwidth}
       \centering
        \includegraphics[width=\textwidth]{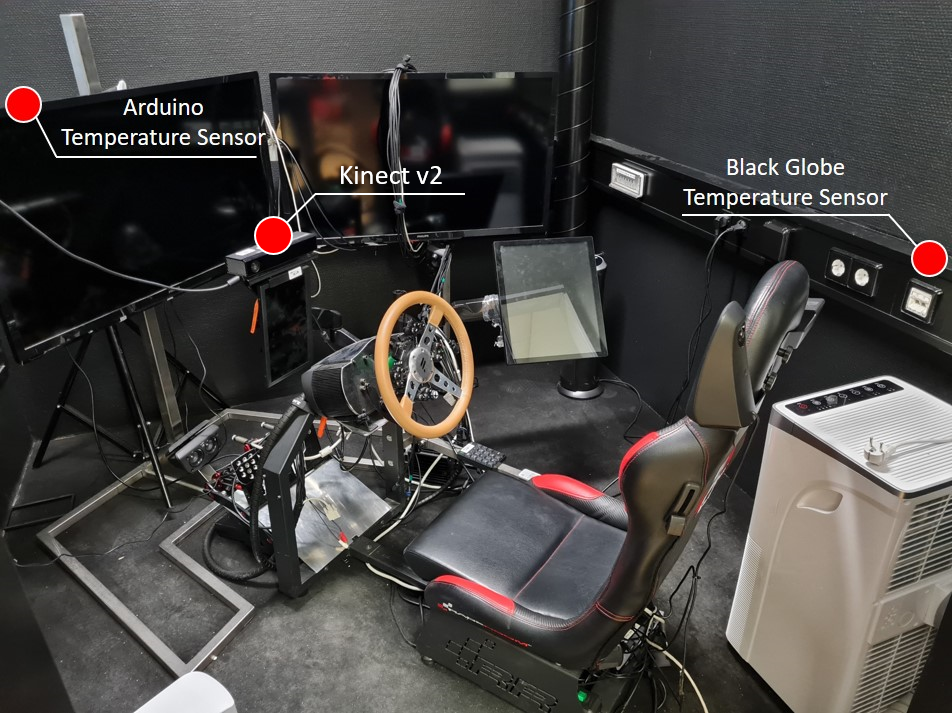}
        \caption{\label{fig:setup_real} The room as used for the thermal comfort recordings. The employed A/C unit can be seen on the bottom right. Temperature sensors were placed above the A/C unit and on top of the monitor on the left-hand side.}
        \Description{This Figure shows the setup that was used for the thermal comfort study in a schematic form.}
        \end{subfigure}\hspace{2mm}
         \begin{subfigure}[c]{0.48\textwidth}
          \centering
                 \includegraphics[width=0.74\textwidth]{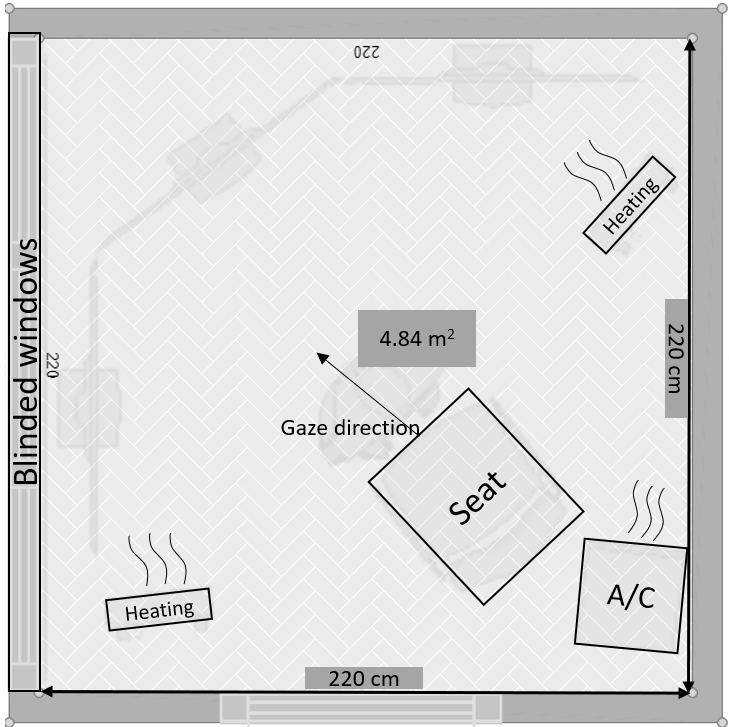}
        \caption{\label{fig:setup_overview} The schematic room plan used for the thermal comfort recordings. A 3d version with stylized devices can be found via this \href{https://roomtodo.com/planner/project/5021113de214}{link}.\newline}
        \Description{This Figure shows the setup that was used for the thermal comfort study. In the top-left corner, three displays are shown. In front of the displays, a steering wheel and a chair can be seen. Behind the chair, an A/C unit is shown.}
         \end{subfigure}
         \caption{Overview of the climate chamber.}
\end{figure*}

\textbf{Sensory setup.} \citet{ciuhaEffectThermalTransience2019} used a climatic chamber and spanned $25$°C (range=[$15$°C, $40$°C]) over $150$ min. Due to the unavailability of such a professional chamber, we built a low-fidelity climatic chamber using mobile A/C devices. The room had a size of 220cm $\times$ 220cm with a height of 280cm (see \autoref{fig:setup_overview}). This results in a volume of 13550l. While previous scarcely report the room volume (e.g., \citet{battistel2023investigation} do not report this), this is in the range of available commercial climate chambers (e.g., see \href{https://clitec.ch/en/fields-of-activity/environmental-simulation/walk-in-climate-chambers/}{Clitec}). The room is painted completely black, from the carpet to the walls and ceiling to the blinds for the 3 small windows. We used the Monzana MZKA1000 Smart A/C (see \autoref{fig:setup_real}) with a power output of $9000$ BTUs as it provides cooling and heating modes and includes a smart home cloud that allows for developer access via APIs. This device enables temperature changes in the range of [$16$°C, $32$°C]. As the A/C unit's heating capability is less powerful than its cooling mode, two additional smart home heating units (Nedis P22-2054875) were added. The smart heating components allow for temperature changes in the range of [$15$°C, $35$°C] and include two heating modes ($1.200$W/$2.000$W). Adding multiple A/C units for heating also ensures that the area is heated at multiple locations, which allows for a more uniform temperature change. Other than the mentioned smart A/C and heating units, no other components were used to manipulate the ambient temperature (i.e., no seat heating). A series of internal tests showed that the ambient temperature range achieved most reliably within $60$ min was [$18.4$°C, $32.0$°C]. The time interval of $60$ min was selected to ensure that a mean temperature change rate of $0.45$°C/min is achieved for both heating and cooling. This change rate is favorable compared to a high rate of $1$°C/min as large thermal changes over short periods are perceived more intensely and, therefore, shift the range at which a person feels comfortable at~\cite{ciuhaEffectThermalTransience2019}.

\begin{table*}[ht!]
\footnotesize
\centering
\caption{\label{tab:inputs} List of derived modalities from different input spaces for multi-modal thermal comfort estimation.}
\begin{tabular}{lp{12.5cm}}
\toprule
\textbf{Input Space} & \textbf{Modalities}\\ 
\midrule
Personal Context & Age~\cite{DELFERRARO2015177, GUERGOVA201180}, Gender~\cite{CHAUDHURI2018391, karjalainenThermalComfortGender2012}, Clothing~\cite{ISO7730, ramspeckASHRAESTANDARDSCOMMITTEE}, Tiredness~\cite{griefahnEffectsGenderAge2001}, Weight \& Height~\cite{Bodymass2020}, Body Fat~\cite{bodyfat2001a}, Metabolic Rate \& Activity Level~\cite{ISO7730, ramspeckASHRAESTANDARDSCOMMITTEE, EffectTemperatureMetabolic2013}\\ [0.2cm]
External Context & Relative Humidity~\cite{ISO7730, jingImpactRelativeHumidity2013}, Ambient Temperature~\cite{ciuhaEffectThermalTransience2019, ramspeckASHRAESTANDARDSCOMMITTEE}, Radiation Temperature~\cite{ISO7730, ATMACA20073210, ramspeckASHRAESTANDARDSCOMMITTEE}, Air Velocity~\cite{ISO7730, ramspeckASHRAESTANDARDSCOMMITTEE}\\ [0.2cm]
Physiology & Heart Rate~\cite{mansiMeasuringHumanPhysiological2021, maoThermalComfortEstimation2021}, Wrist Skin and Body Core Temperature~\cite{CHAUDHURI2018391, InfluenceOfBodyTemp1999, YAO2008310}, Galvanic Skin Response (GSR)~\cite{mansiMeasuringHumanPhysiological2021}\\ [0.2cm]
Visual Attributes & 3D Body Pose~\cite{yangRealtimeContactlessMeasurements2019,YANG2020110261}, RGB View~\cite{zhangFrownBasedThermalComfort2021} \\[0.2cm]
Emotion & Emotional States (after \citet{ekman})~\cite{WANG2020109789} with neutral emotion\\
\bottomrule
\end{tabular}
\end{table*}

The PCE-WB 20SD thermometer was selected due to its logging rate of $1$Hz and the ability to record ambient temperature, relative humidity, and radiation temperature with the integrated black globe components~\cite{pce}. Black globe temperature is measured with the thermometer being inside a black globe. This means that this indicates how hot it feels in direct sunlight. A cheaper and easily integrable solution for temperature and humidity data is Arduino sensory units~\cite{arduino}. However, these kits do not provide the same level of accuracy as specialized measuring tools do. Nevertheless, ambient temperature and humidity data from an Arduino sensory board were included in the data logging application to be able to compare prediction performance with different frequencies and levels of accuracy for temperature and humidity streams. As for physiological signals, we used the Empatica E4~\cite{empatica}. Modalities like emotion, body pose, and visual features can be captured using appropriate machine-learning models and RGB frame processing. For emotion estimation, \textit{Serengil and Ozpinar} presented a deep learning model that includes multiple face detector backends and allows for analysis of recorded RGB frames~\cite{serengil2020lightface, serengil2021lightface}. For 2D pose key point estimation, we employed OpenPose~\cite{openpose}. Seeing as both models can be executed in real-time, they were integrated for emotion and body pose estimation. RGB frames themselves can easily be captured using a webcam. However, for the data collection process, Microsoft's Kinect v2 for Windows~\cite{kinectDocs} was selected as it provides RGB frames of size $1920\times1080$ and depth frames of size $512\times424$~\cite{kinect}. As the selected body pose model estimates 2D skeleton key points, the depth frame provided by the Kinect sensor is necessary to measure the respective depth values. This way, 2D key points are extended to 3D key points (see \autoref{tab:inputs}).

\subsection{Participants}
We gathered data from \N{21} participants. The gender ratio was \textit{12F/9M} with a mean age of \m{24.64} (\sd{3.03}, \rg{20}{33} in years). Participants weighed \m{69.97} (\sd{15.02}, \rg{53.00}{106.90} in kg) at an average height of \m{174.50} (\sd{10.18}, \rg{155.00}{198.00} in cm) and a body fat percentage of \m{22.00\%} (\sd{5.00\%}, \rg{14.00\%}{34.00\%}). Body temperature measured at the forehead with an infrared skin thermometer was mostly similar for all participants (\m{36.38}, \sd{0.31}, \rg{35.8}{37.2} in °C). Clothing levels varied only slightly, as most participants wore a short-sleeve T-shirt and trousers (mean insulation based on ASHRAE standard 55 clothing insulation tables \cite{ramspeckASHRAESTANDARDSCOMMITTEE} \m{0.60}, \sd{0.05}, \rg{0.45}{0.69} in clo).
Only 14\% (3Y/18N) of participants reported having performed physical activities in the hour before the recording session, while the time since participants' last meals was \m{4.23} hours (\sd{4.92}, \rg{0}{20}) before the recording. Based on the tiredness ratings in the initial personal context assessment, participants reported having been moderately tired at the start of their recording (\m{4.12}, \sd{1.73}, \rg{2}{8}).


\begin{table}[ht!]
\centering
\footnotesize
\caption{\label{tab:dataset-statistics} Comparison of minimal and maximal values, means and standard deviations of the numeric features from the \indoor and \vehicleDataset. \textit{ibp} stands for the (averaged) image brightness per pixel as an approximation for solar radiation.}
\begin{tabular}{lccccc}
\toprule
                                 &     &   \multicolumn{2}{c}{\indoorDataset} & \multicolumn{2}{c}{\vehicleDataset} \\
\midrule
Participants                     &     &   \multicolumn{2}{c}{18}             & \multicolumn{2}{c}{20}              \\
Duration of recordings           & minutes &   \multicolumn{2}{c}{60}             & \multicolumn{2}{c}{30}              \\
                                 &     &   \minmax{black}{black}{min}{max}     & \meanstd{mean}{std}     &  \minmax{black}{black}{min}{max}   & \meanstd{mean}{std}         \\
\midrule 
Radiation Temperature            & °C  &   \minmax{black}{black}{16.9}{33.6}   & \meanstd{25.53}{3.92}   &  \multicolumn{2}{c}{-}                                           \\
Heart Rate                       & bpm &   \minmax{black}{black}{40.0}{191.99} & \meanstd{82.94}{13.46}  &  \minmax{black}{black}{37.28}{191.99} &  \meanstd{77.70}{24.74}  \\
Wrist Skin Temperature           & °C  &   \minmax{black}{black}{27.91}{36.95} & \meanstd{33.82}{1.75}   &  \minmax{black}{black}{25.55}{36.43}  &  \meanstd{30.89}{2.54}   \\
Galvanic Skin Response           & $\mu$S~(microsiemens) & \minmax{black}{black}{0.0}{16.9}    & \meanstd{1.42}{2.81}    &  \minmax{black}{black}{0.01}{9.19}    &  \meanstd{0.57}{1.21}    \\
Ambient Temperature PCE-WB 20SD  & °C  &   \minmax{black}{black}{17.1}{33.7}   & \meanstd{25.31}{3.72}   &  \multicolumn{2}{c}{-}                                           \\
Ambient Temperature Arduino      & °C  &   \minmax{black}{black}{17.6}{37.0}   & \meanstd{26.82}{4.52}   &  \minmax{black}{black}{10.0}{35.40}   &  \meanstd{25.13}{4.23}   \\
Relative Humidity                & \%  &   \minmax{black}{black}{12.0}{55.0}   & \meanstd{31.35}{8.86}   &  \minmax{black}{black}{10.0}{67.0}    &  \meanstd{29.54}{9.20}   \\
Solar Radiation                  & ibp &   \minmax{black}{black}{0.16}{0.45}   & \meanstd{0.28}{0.04}    &  \minmax{black}{black}{0.0}{0.88}     &  \meanstd{0.46}{0.06}    \\
\bottomrule
\end{tabular}
\end{table}

\subsection{Raw Dataset}\label{sec:raw_dataset}
\begin{figure}[ht]
    \centering
        \includegraphics[width=0.75\textwidth]{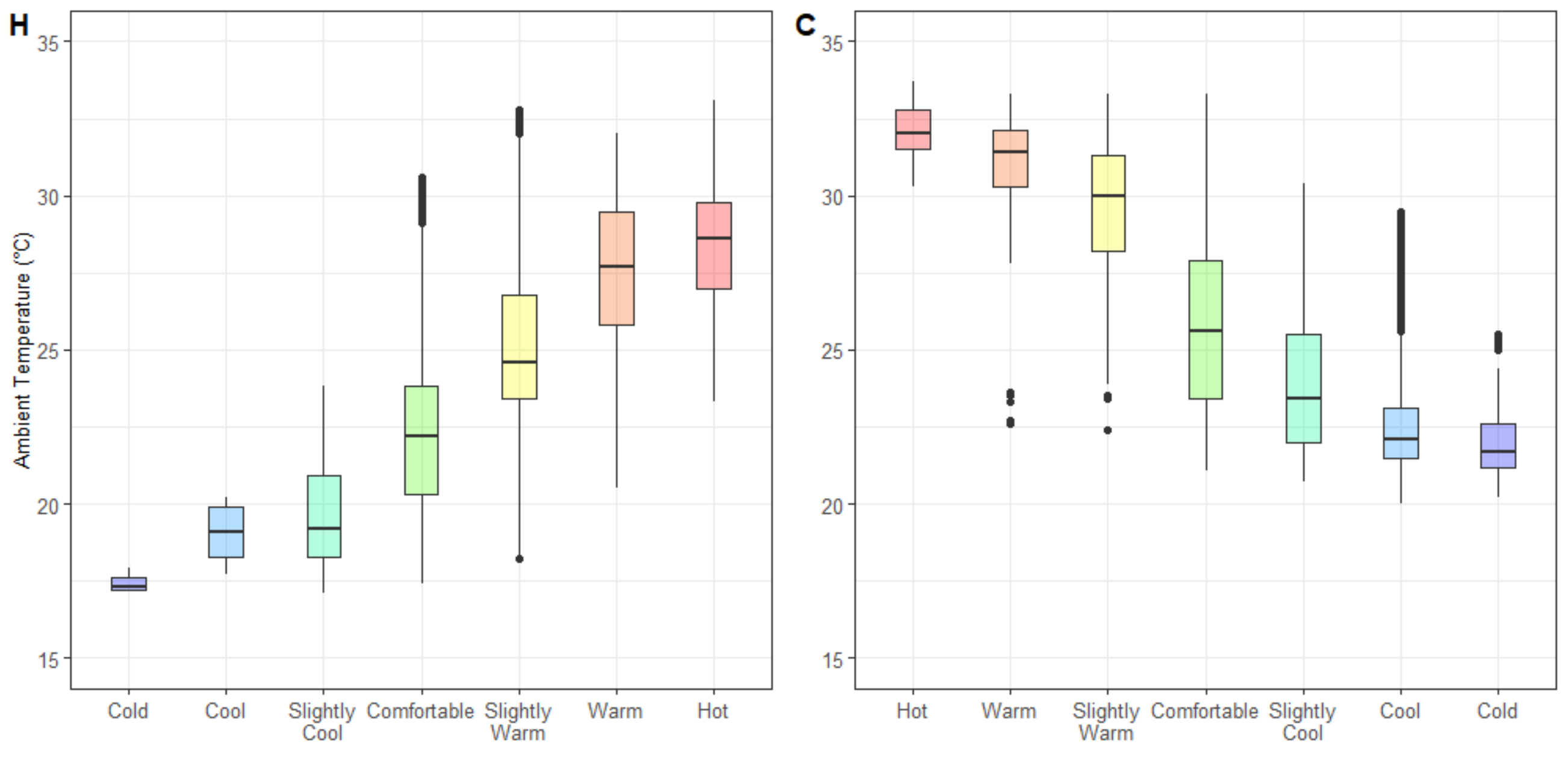}
        \caption{\label{fig:hot_cold} Left: Labels given during heating. Right: Labels for the cooling phase. In line with previous research (e.g.,~\cite{ciuhaEffectThermalTransience2019}), thermal comfort is established at a lower ambient temperature during heating than during cooling.}
        \Description{This Figure shows two visualizations for labels given during different temperatures in the heating (left side) and cooling phase (right side). On the left side, the plot illustrates that only a few participants rated the achieved ambient temperatures as cold, and most participants rated achieved comfortably as hot states. On the right side, the plot illustrates that all possible thermal comfort states were generally achieved at higher temperatures. It can also be seen that cold and cool thermal comfort states were not achieved by most participants.}
\end{figure}
The initial raw dataset included recordings from all $21$ participants. Due to incorrect use of the numeric keyboard during labeling, the data of one participant was removed. Also, we removed another $2$ participants due to incomplete radiation temperature readings. Therefore, the final and filtered raw dataset includes $18$ participants and a separate image archive with the RGB frames. We used $16$ participants for training and $2$ participants to evaluate our models (see Section~\ref{sec:experiments}). The dataset includes corresponding body pose key-point coordinates for each recording. Due to movement out of the depth camera's field of view during the recordings, not all key points could be estimated reliably at all times, which led to empty vectors in the dataset. 

The final dataset includes a total of $1.856.290$ data points with extrapolated labels, each with $34$ feature columns (including timestamp and label columns) from which $2927$ data points were actually labeled by the user. \autoref{tab:dataset-statistics} shows for the \indoorDataset a descriptive evaluation for the numeric features not gathered using the self-report GUI. \textit{Radiation and ambient temperature} are fairly similar and could replace each other during classification. Concerning the values measured with the Empatica E4 wristband, the most stable measurements were achieved for the wrist skin temperature. This is indicated by the low standard deviation (\sd{1.75}) and minimum and maximum values (\textit{min=27.81}, \textit{max=36.95}) close to the mean of $33.82$°C. Contrarily, the raw measurements for the remaining physiological signals, heart rate, and GSR were far less stable. The highest outlier rate out of the numeric continuous features was found in the GSR measurements ($13.7\%$ outliers). Therefore, a data pre-processing scheme that includes outlier removal methods is required when using the dataset for classifier training. 
The PMV calculation also requires the air velocity caused by body movement and not the air velocity measured by a given air velocity sensor. Thus, an approximation was used.
Air velocity was not measured but deduced from the capabilities of the A/C unit as listed in the \href{https://www.manualslib.com/products/Monzana-Mzka1000-12111205.html}{manual}. The manual states that a max air volume flow of 360 m/h is possible (page 37). Therefore, we set the air velocity to static 0.1 m/s. 

The raw dataset further includes two \textit{emotion} feature columns, as participants were able to report their current emotions (after \citet{ekman} with the additional \textit{neutral} emotion; exclusive choice: ``Which item describes your current emotion best?'' with the options: Anger, Fear, Sadness, Disgust, Happiness, Neutral, Surprise, and Contempt; see \autoref{sec:introduction}) using the logging GUI while also having their emotions estimated using the RGB capture and the DeepFace model. Emotions were estimated at the same frequency as emotion self-report dialogues were displayed. Most participants rated their emotions as \textit{neutral}. This resulted in different rating distributions across emotion feature columns. Self-reported emotions were distributed as follows: \textit{Anger 0.83\%, Contempt 0.00\%, Disgust 3.14\%, Fear 0.00\%, Happiness 8.35\%, Neutral 86.38\%, Sadness 0.73\% and Surprise 0.49\%}. The model-based emotion predictions also tended towards neutrality but less strongly: \textit{Anger 10.99\%, Disgust 0.08\%, Fear 13.48\%, Happiness 8.98\%, Neutral 41.14\%, Sadness 20.87\%, and Surprise 4.46\%}. This indicates a mismatch between felt emotions and detected facial expressions by DeepFace. Therefore, we refrained from using the data from the DeepFace model. We also did not use the self-report emotion data due to the very low variance. The dataset also includes two different \textit{ambient temperature} features. As described at the beginning of this section, an external thermometer was used during the recordings to measure radiation temperature changes. However, when comparing the ambient temperature measures of the external thermometer and the Arduino sensory kit measures, it could be observed that the Arduino kit tends to react more intensely when ambient temperature changes occur. Moreover, the temperature sensor used in the sensory kit is labeled to have an accuracy of $2$°C \cite{arduinodht11}, whereas the external thermometer (PCE-WB 20SD) is labeled with an accuracy of $0.8$°C \cite{pcelogger}. For this reason, further analysis and visualizations of ambient temperatures are based on the values measured with the more accurate PCE thermometer. The label distribution throughout the dataset suggests that the minimum temperature during the trials was insufficient for inducing cold and cool thermal sensations. The answers given during participants' debriefing further support this assumption, as it was mentioned that \textit{"the temperature, in the beginning, felt somewhat cool, but you get used to it quickly"} [P16] and \textit{"it didn't get very cold, but it did get quite hot."} [P7]. The collected labels are distributed as follows: \textit{Cold 4.49\%, Cool 9.53\%, Slightly Cool 22.27\%, Comfortable 22.71\% (this would represent the accuracy of a null model predicting only ``comfortable''), Slightly Warm 13.30\%, Warm 15.83\%, Hot 11.88\%}. A difference in thermal comfort ratings could be found between the heating and cooling phases, as illustrated in \autoref{fig:hot_cold}. This is in line with previous findings that reported a thermal comfort zone shift based on previous exposure to different thermal environments~\cite{ciuhaEffectThermalTransience2019}. Moreover, a thermal comfort zone can be seen between the labels \textit{Slightly Cool} and \textit{Slightly Warm}, as the temperature ranges for slightly cool, comfortable, and slightly warm states are the largest among all reported states. From a classification perspective, this indicates that classifications of states in the thermal comfort zone may be more difficult to predict based on ambient temperature alone, while colder states may be more difficult to predict due to the imbalance in frequency of occurrence in the dataset.\\
\textbf{While some of these features might not encode much information due to inherent limitations (e.g., 3D pose limited by the available space), we emphasize that additional data might facilitate novel approaches. Therefore, we strived to collect a rich and large-scale dataset.}

\section{Thermal Comfort Vehicle Dataset}\label{sec:vehicle_dataset}
\begin{figure}[ht]
    \centering
        \includegraphics[width=0.5\textwidth]{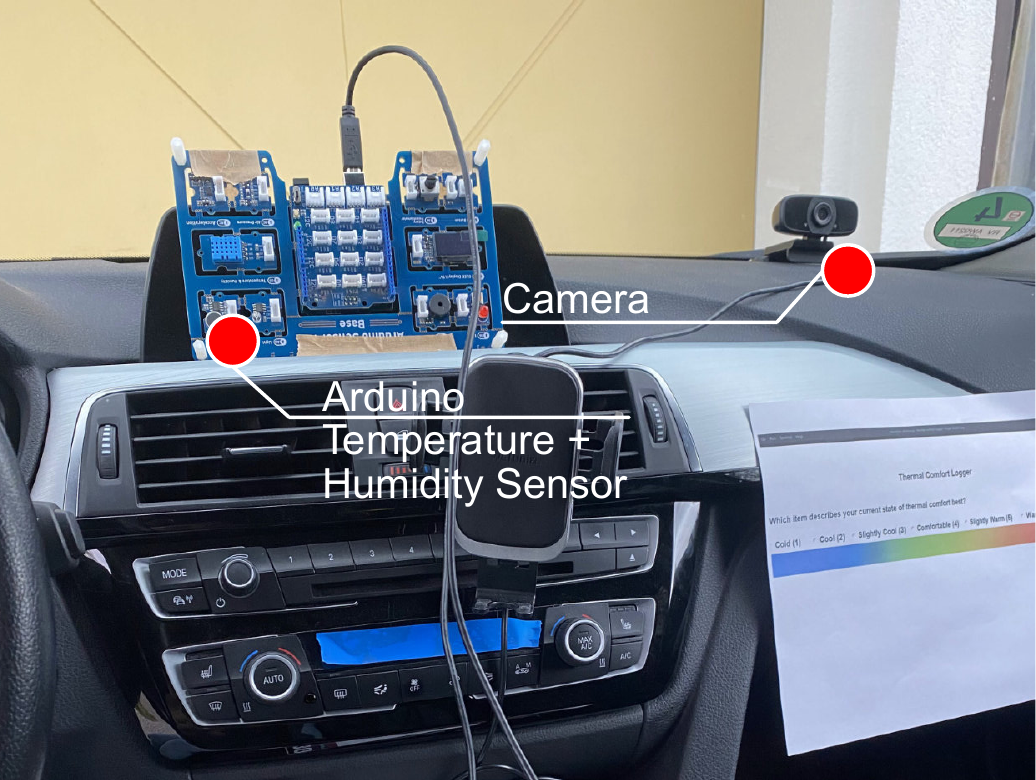}
        \caption{\label{fig:setup_vehicle} Sensory setup inside the BMW $3$ F31. The Arduino board was mounted on top of the middle A/C vents. The displays were covered to hide temperature information from participants. In-vehicle setup did not include a Kinect camera.}
        \Description{This Figure shows the setup that was used inside the car for the thermal comfort study. In the top region, the Arduino board is shown covering the display. Below the Arduino board, the A/C vents are shown. Further below, a second display is covered using tape.}
\end{figure}

\begin{figure}[ht]
    \centering
        \includegraphics[width=0.6\textwidth]{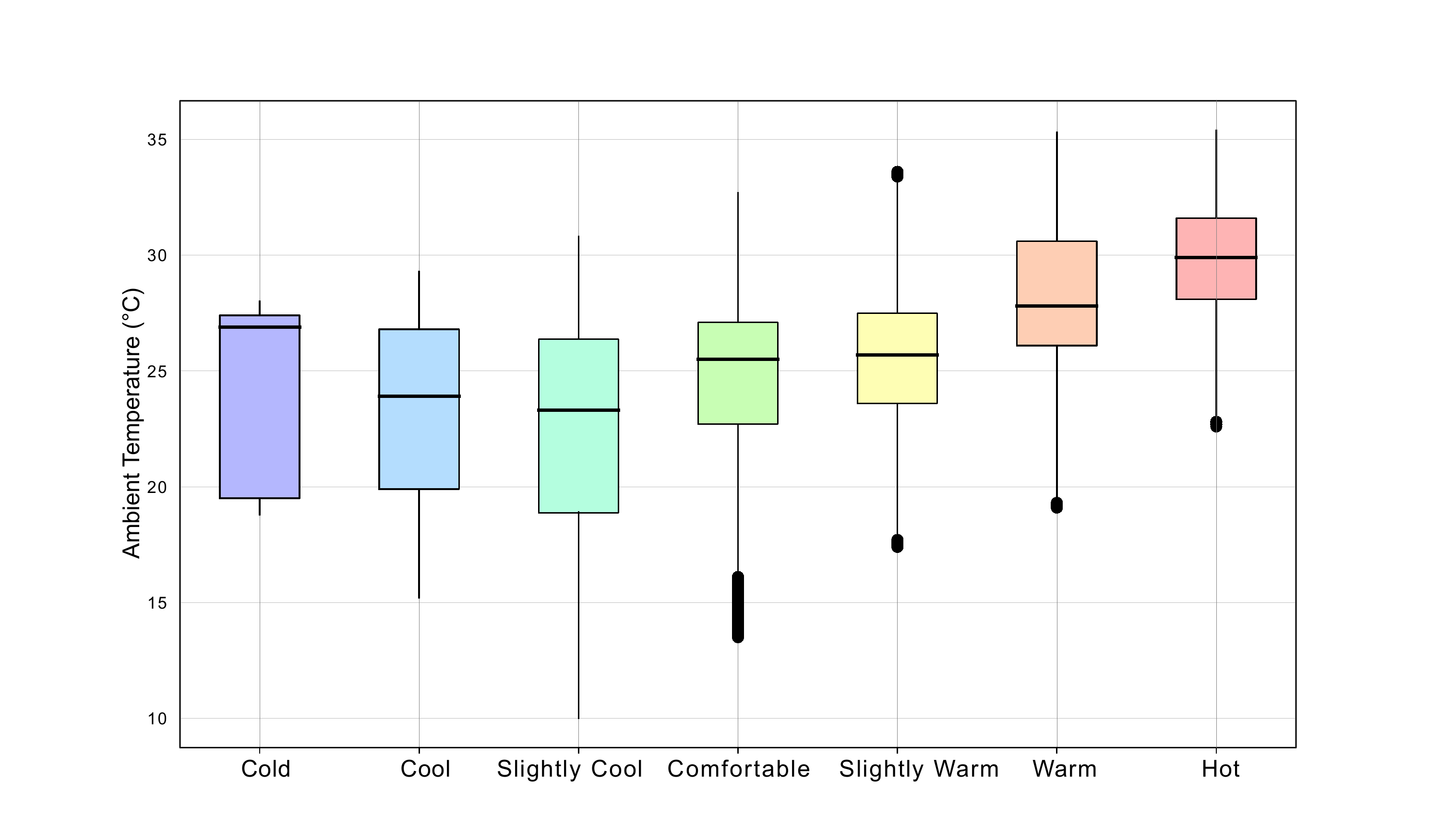}
        \caption{\label{fig:real_temp} Labels given during in-vehicle study.}
        \Description{This Figure shows two visualizations for labels given during different temperatures in the heating (left side) and cooling phase (right side). On the left side, the plot illustrates that only a few participants rated the achieved ambient temperatures as cold, and most participants rated achieved comfortably as hot states. On the right side, the plot illustrates that all possible thermal comfort states were generally achieved at higher temperatures. It can also be seen that cold and cool thermal comfort states were not achieved by most participants.}
\end{figure}

To collect data for a real vehicle scenario, we used a BMW $3$ F31 for conducting a second study and data acquisition (see \autoref{fig:setup_vehicle}). We refer to this dataset as \textbf{\vehicleDataset}. During the study, we utilized the application described in Section~\ref{sec:thermal_comfort_study}. To manipulate temperature, we used the built-in A/C and adjusted the temperature manually. The temperature range was narrower compared to the \indoor study due to the lower power of the built-in A/C. Participants provided data annotations in a time frame of $30$ min. $14$ participants provided data while the vehicle was parked, $6$ while the experimenters drove the vehicle in a small town and highways in the region of \anon{Bad Waldsee, Germany}. The participants remained in the passenger seat. The possible inputs were shown on a printout (see \autoref{fig:setup_vehicle}). During the in-vehicle study, users were notified by a beep sound to provide a rating of their current thermal comfort state. We determined the input spaces and modalities shown in \autoref{tab:inputs} such that the PMV variables (ambient temperature, relative humidity, metabolic rate, clothing insulation, radiation temperature, air velocity)~\cite{fanger1970thermal} along with modalities for the identified physiological, visual, emotional, and personal input spaces are included. Additionally, we approximate solar radiation by computing the average brightness of an image frame extracted from the RGB video that was captured during the study. 

\subsection{Participants}
We gathered data from \N{20} participants, two of whom had participated in the \indoor study. The gender ratio was \textit{7F/13M} with a mean age of \m{33.41} (\sd{12.67}, \rg{25}{66} in years). Participants weighed \m{73.52} (\sd{23.47}, \rg{53.00}{119.0} in kg) at an average height of \m{179.89} (\sd{10.85}, \rg{161.00}{198.00} in cm). Body temperature measured at the forehead with an infrared skin thermometer was mostly similar for all participants (\m{36.14}, \sd{0.15}, \rg{35.8}{36.4} in °C). Clothing levels varied only slightly, as most participants wore a long-sleeve shirt and trousers (mean insulation based on ASHRAE standard 55 clothing insulation tables \cite{ramspeckASHRAESTANDARDSCOMMITTEE} \m{0.67}, \sd{0.06}, \rg{0.57}{0.81}).
None of the participants reported having performed physical activities in the hour before the recording session, while the time since participants' last meals was \m{3.75} hours (\sd{3.83}, \rg{0}{14}) before the recording. Based on the tiredness ratings in the initial personal context assessment, participants reported having been moderately tired at the start of their recording (\m{2.75}, \sd{1.13}, \rg{1}{5}).


\subsection{Raw Dataset}\label{sec:raw_dataset_in_vehicle}
To investigate the thermal comfort of humans under real conditions, we collected a data set using a real car (i.e., a BMW 3). The study was carried out from the end of March until the end of April 2023, with varying outdoor temperatures ranging from $6$ to $14$°C. During the study, sensory data of the Empatica E4 for wrist skin temperature, GSR, and heart rate were recorded. Additionally, we recorded ambient relative humidity and ambient temperature using an Arduino board. We did not measure air velocity in the vehicle due to the necessity to report the velocity in relation to the participants' movement. We report the mean, standard deviation, and minimal and maximal value for the corresponding sensory readings in \autoref{tab:dataset-statistics}. For the study, we randomized the following conditions, which were each conducted one or multiple times in random order: switching off the A/C, switching to maximum heating ($28$ °C), switching to maximum cooling ($16$ °C), opening the window for at least $1$ min. Additionally, we randomized driving the car and standstill of the car per participant for the study. The collected data set consists of $1.069.374$ labels, which we extrapolate from $1597$ collected labels from the user. 

\section{Comparison to Publicly Available Datasets}\label{sec:datasets_comparison}

We compare our \indoor and \vehicleDataset to related available datasets. 
\autoref{tab:comparison_1} shows that our \tool \indoor dataset has over 17 times the number of entries compared to the next largest dataset from prior work. 
All datasets include measures from the personal and environmental input space. The environmental variables for the \indoorDataset, ASHRAE RP-884, ASHRAE II, and US Office Buildings datasets include the environmental measures necessary for PMV calculation. The Scales Project dataset includes weather station and indoor environment data; however, no specialized tools (e.g., black globe thermometers for radiation temperature measures or anemometers for air velocity) were used. Thus, PMV estimations cannot be computed using the Scales Project dataset. The UCLIC-Bentley Comfort (UBComfort) dataset is the closest dataset to the \tool dataset but contains fewer data points with fewer features. However, it does include additional features such as mental relaxation level and sitting discomfort.
The listed publicly available datasets include ratings that were gathered with different rating scales (\textit{TS=thermal sensation}, \textit{TP=thermal preference}, and \textit{TA=thermal acceptability}). The \tool dataset differs from the publicly available datasets by including physiology, emotion, visual signals, and participants' RGB recordings. Additionally, the \tool dataset includes \textbf{temporal data} compared to the singular data of the other datasets. In \autoref{appendix:data_prep}, \autoref{fig:person-in-lab-in-vehicle} shows an example image (blurred) for the \indoor and \vehicleDataset.


\begin{table}[ht]
\centering
\footnotesize
\caption{\label{tab:comparison_1} Comparison to publicly available datasets with our \toolData. 7P=seven-point thermal sensation (\textit{Cold, Cool, Slightly Cool, Comfortable, Slightly Warm, Warm, Hot}), 3P=three-point thermal preference (\textit{Want Warmer, No Change, Want Cooler}), and 2P=two-point thermal comfort (\textit{Acceptable, Unacceptable}). Measurements coded as P=Personal Context, EX=External Factors, PH=Physiological Factors, E=Emotions, and V=Visual Attributes. The ASHRAE datasets also include different environmental indices (e.g., PMV index).}
\begin{tabular}{@{}lccccccccccc@{}} 
\toprule
\multirow{2}{*}{Dataset} & \multirow{2}{*}{Participants} & \multirow{2}{*}{Entries} & \multirow{2}{*}{Temporal Data} & \multicolumn{5}{c}{Included Measurements} & \multicolumn{3}{c}{Rating Scales} \\
\cmidrule(lr){5-9}\cmidrule(lr){10-12}                   
{}                                   &     {}  & {} & {}   & P          & EX         & PH        & E         & V         & 7P         & 3P         & 2P         \\
\midrule
ASHRAE RP-884~\cite{ashrae1}     &     $25.288$              & $25.288$ & \redxmark         &\greencheck &\greencheck & \redxmark & \redxmark & \redxmark &\greencheck &\greencheck &\greencheck  \\
ASHRAE II~\cite{parkinson2022ashrae} & $109.033$ & $109.033$   & \redxmark      &\greencheck &\greencheck & \redxmark & \redxmark & \redxmark &\greencheck &\greencheck &\greencheck  \\
Scales Project~\cite{scales_project}  &  $8.225$                    & $8.225$  & \redxmark         &\greencheck &\greencheck & \redxmark & \redxmark & \redxmark &\greencheck &\greencheck &\greencheck  \\
US Office Buildings~\cite{us_office} &      $24$            & $2.503$  & \redxmark         &\greencheck &\greencheck & \redxmark & \redxmark & \redxmark &\greencheck &\greencheck &\greencheck  \\
OccuTherm~\cite{10.1145/3360322.3360858}    &    $77$                    & $2.067$   & \greencheck           &\greencheck &\greencheck & \greencheck & \redxmark & \redxmark &\redxmark &\redxmark &\redxmark  \\
LPTC~\cite{10.1145/3485730.3493693}       &      $17$                    & $1.403$   & \greencheck           &\greencheck &\greencheck & \greencheck & \greencheck & \redxmark &\redxmark &\greencheck &\redxmark  \\

UBComfort~\cite{9597393}       &      $28$                    & $587$   & \greencheck           &\greencheck &\greencheck & \greencheck & \redxmark & \greencheck &\greencheck& deducible  & deducible  \\

\hdashline
\textbf{\tool \indoor}         & $18$                &\textbf{$1.856.290 (2.927^*)$}& \greencheck&\greencheck &\greencheck &\greencheck&\greencheck&\greencheck&\greencheck& deducible  & deducible  \\
\textbf{\tool \vehicle}        & $20$                &\textbf{$1.069.374 (1.597^*)$}& \greencheck&\greencheck &\greencheck &\greencheck&\greencheck&\greencheck&\greencheck& deducible  & deducible  \\
\bottomrule
\multicolumn{12}{l}{*Number of original labels provided by the participants.}
\end{tabular}
\end{table}

\section{Validation}\label{sec:experiments}
We conducted several experiments on the \tool \indoor dataset acquired (see Section~\ref{sec:indoor_dataset}). First, we performed feature importance ranking using an impurity-based method. Based on the found feature importance, we conducted a feature combination study in Section~\ref{ssec:feature_combination_study} to find the best combination of features. In \autoref{sec:feature-permutation-importance}, we investigate feature importance of physical, psychological, and physiological features by performing feature permutation based on RF and LSTM model estimations. Having these insights, we considered an input vector of four features for all experiments and implemented three types of classification models. As our baseline, we use an RF classifier (Section~\ref{ssec:random_forest_classifier}). We also measure the performance of the PMV index on our dataset in Section~\ref{ssec:pmv_performance}.
In Section~\ref{ssec:exp_comparison}, after the evaluation of the best classifier, we continue with a comparison of our dataset to the ASHRAE II~\cite{parkinson2022ashrae} dataset. 
We compare these against deep learning models based on recurrent networks (RNN) (Section~\ref{ssec:LSTM_classifier}), and a combination of RNN and convolutional neural networks (CNN), in Section~\ref{ssec:RCNN_classifier}. Additionally, we investigate the forecasting of time series data to predict thermal comfort for a future state in Section~\ref{ssec:forecasting}. Next, we provide a short description of the used metrics to quantify estimation accuracy.\\

\textbf{Thermal Comfort Metric.} To measure a classifier's performance, we use three metrics of accuracy with varying scales of precision: $\kappa_n$, where $n$ is the number of classes used to compute accuracy. First, the seven-point thermal sensation scale (\textit{Cold, Cool, Slightly Cool, Comfortable, Slightly Warm, Warm, Hot}), which is denoted as \seven. Second, the three-point thermal preference scale (\textit{Cold, Comfortable, Warm}), denoted as \three. For computing \three, we reduce seven classes by mapping the classes (\textit{Cold, Cool, Slightly Cool}) to \textit{Cold}, (\textit{Slightly Warm, Warm, Hot}) to \textit{Warm}, and \textit{Comfortable} remains.
And third, the two-point thermal comfort scale (\textit{Comfortable, Uncomfortable}), which we denote as \two. Note that for computing \two, we reduce seven classes to binary classes by mapping the classes (\textit{Slightly Cool, Comfortable, Slightly Warm}) to \textit{Comfortable} and the remaining classes to \textit{Uncomfortable}.\\

For all experiments, our training dataset consists of data from 16 participants, while our test data consists of the two remaining participants. For the LSTM approach (the \textbf{best-performing approach}, see Section~\ref{ssec:LSTM_classifier}), we additionally conducted a 20-fold cross-validation (see \autoref{sec:cross-validation}), showing robust performance. 

\tool represents the only approach to have acquired both in a field and an in-lab study and to focus on buildings and vehicles. Other work only focused on field studies with a focus on buildings.

\begin{figure*}[ht]
    \centering
        \includegraphics[width=\textwidth]{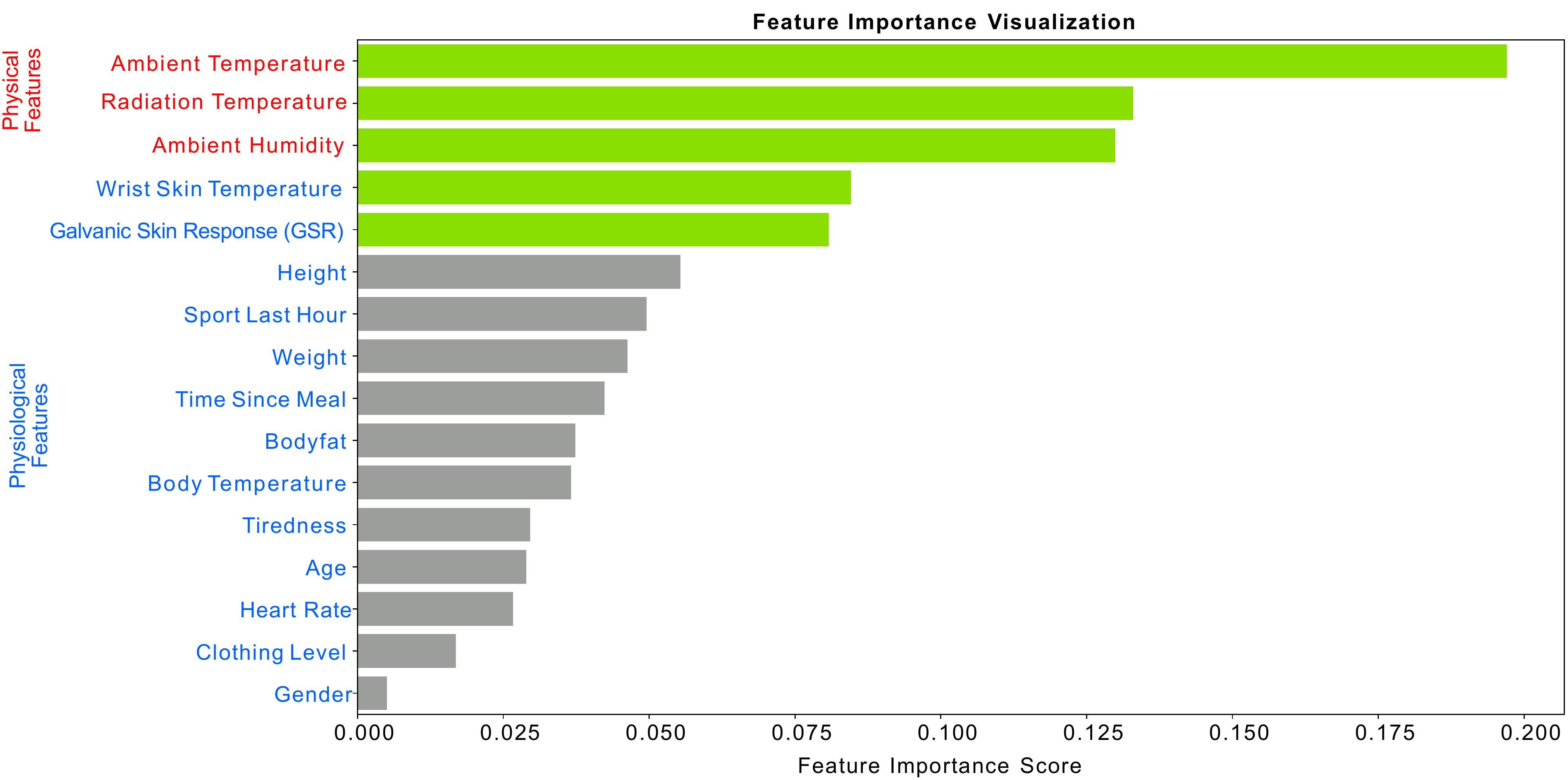}
        \caption{\label{fig:impurity_imp} Feature importance ranking based on mean accuracy impurity decrease of an RF classifier. According to this metric, the five most important features were ambient temperature, relative humidity, radiation temperature, skin temperature, and galvanic skin response, marked in green.}
        \Description{This Figure shows the feature importance ranking based on Random Forest estimated data impurity scores. According to this metric, the five most important features were ambient temperature, relative humidity, radiation temperature, skin temperature, and galvanic skin response. The remaining features were ranked as far less important. According to this metric, the least important feature was the participants' gender.}
\end{figure*}

\subsection{Random Forest Classifier}\label{ssec:random_forest_classifier}
In a first step, we leverage an RF classifier to perform an impurity-based importance ranking of all available features, see \autoref{fig:impurity_imp}. For this task, we use the RF implementation from the \texttt{sklearn} library~\cite{scikit}, which provides functions for dataset sampling, pre-processing, and training pipeline definition. While RF models provided by \texttt{sklearn} allow for tuning of different parameters such as the number of estimators, maximum tree depth, and maximum number of features to consider per node, the standard configuration is used at first and then later optimized using a grid search approach. For our experiments, we used $400$ estimators and a maximum tree depth of $8$ adopting the Gini impurity cost function~\cite{jost2006entropy}. We used every \nth{100} data point to downsample the dataset.\\
\noindent\textbf{Results.} In \autoref{tab:performances}, we report the classification accuracy of \seven$=47.1\%$, \three$=73.8\%$, and \two$=67.9\%$, for the RF classifier, indicating a mediocre performance.

\subsection{PMV and Scale Reduction Performance}\label{ssec:pmv_performance}
To compare the performance of our proposed models to the PMV index, we pre-computed the PMV values for all participants in the evaluation split of our dataset using the clothing level, radiation temperature, ambient temperature, and relative humidity features. For computation of PMV measures, the \texttt{PyThermalComfort} package~\cite{tartariniPythermalcomfortPythonPackage2020a} was used, which enables PMV computation based on the definitions in ISO7730~\cite{ISO7730} and ASHRAE standards~\cite{ramspeckASHRAESTANDARDSCOMMITTEE}. \\
\noindent\textbf{Results.} A comparison with the participants' subjective thermal sensation ratings showed that the PMV index fails to accurately predict thermal comfort. The achieved prediction accuracies are \seven$=35.9\%$, \three$=63.7\%$, and \two$=65.2\%$, suggesting that the PMV index values were mostly off by one class.
\begin{table}[ht]
\setlength{\tabcolsep}{3.5pt}
\centering
\footnotesize
\caption{\label{tab:performances} Evaluation results of a performance comparison between different classifiers on the \indoorDataset. In the last row, we report the performance of the PMV index computed for the corresponding PMV values. }
\begin{tabular}{lccc}
\toprule
\textbf{Classifier} & \seven & \three & \two \\
\midrule
LSTM               & $59.3\%$ & $83.5\%$ & $71.9\%$ \\
Forecast 10 sec    & $56.6\%$ & $80.4\%$ & $74.4\%$ \\
Forecast 5 min     & $55.5\%$ & $77.7\%$ & $74.6\%$ \\
Forecast 10 min    & $50.9\%$ & $77.1\%$ & $73.3\%$ \\
CNN-LSTM           & $48.5\%$ & $69.4\%$ & $76.0\%$ \\
Random Forest      & $47.1\%$ & $73.8\%$ & $67.9\%$ \\
\hdashline
PMV                & $35.9\%$ & $63.7\%$ & $65.2\%$   \\
\bottomrule                                             
\end{tabular}
\end{table}

\subsection{ASHRAE Thermal Comfort Field Measurements}
\label{ssec:exp_comparison}
In this experiment, we focus on a comparison of our \indoorDataset to existing thermal comfort datasets. After a close inspection of publicly available datasets, the ASHRAE II~\cite{parkinson2022ashrae} dataset features comparable aspects to our collected dataset, as it provides measurements for \textit{Radiation Temperature}, \textit{Ambient Temperature}, and \textit{Relative Humidity}, which are also measured in our dataset. While our dataset provides sequences over time, ASHRAE II recorded single data points, which makes it unfeasible to compare against our recurrent-based methods. Instead, we optimize an RF classifier, described in Section~\ref{ssec:random_forest_classifier}, using the ASHRAE II dataset. After filtering out data points with incomplete measurements, a total of $31.500$ data points remain. We use $80\%$ of the ASHRAE II data for training, resulting in $25.204$ data points and the remaining $6.301$ data points for evaluation. Further, we use the trained classifier from Section~\ref{ssec:random_forest_classifier}, which is trained on our dataset, and evaluate it on ASHRAE II. Finally, we use the classifier trained on ASHRAE II and evaluate it on our dataset. We report evaluation results in \autoref{tab:ashrae_comparison}. \\
\noindent\textbf{Results.} The evaluation split of our dataset seems to be more difficult than ASHRAE II. This is indicated by a worse performance of the classifier trained and evaluated on our dataset than the classifier trained and evaluated on ASHRAE II. Also, the performance of the classifier trained on ASHRAE II and evaluated on our dataset performs worse than the classifier trained on our data and evaluated on ASHRAE II, which further supports our observation. This comparison shows a big gap between both thermal comfort datasets while underlining the missing feature information when only three sensor measurements are used. Further, this experiment shows a drastic performance drop when only single thermal state values are used for state recognition. Looking at the 2-point performance measure, the classifier trained on our dataset generalizes to the ASHRAE II dataset.

\begin{table}[ht]
\setlength{\tabcolsep}{3.5pt}
\centering
\footnotesize
\caption{Comparison results of an RF classifier that was optimized on our \indoorDataset (\tool) from the climatic chamber and evaluated on the ASHRAE II dataset and vice versa. We report evaluation performance for three metrics: \seven, \three, \two. Each classifier was trained using \textit{Radiation Temperature}, \textit{Ambient Temperature}, and \textit{Relative Humidity} as input vectors.}\label{tab:ashrae_comparison}
\begin{tabular}{ccccc}
\toprule
Train & Evaluate & \seven & \three & \two \\
\midrule
\tool       & \tool       & 40.6\% & 76.1\% & 69.1\% \\
ASHRAE II & ASHRAE II & 45.8\% & 45.8\% & 82.7\% \\
\tool       & ASHRAE II & 16.2\% & 22.3\% & 82.8\% \\
ASHRAE II & \tool       &  2.7\% & 39.7\% & 37.5\% \\
\bottomrule                                             
\end{tabular}%

\end{table}

\subsection{Classification Using Time-Dependent Information}\label{ssec:LSTM_classifier}
We investigated representing our time-dependent data as a time series of feature vectors and leveraging a recurrent neural network to process sequences of features. Thus, we implement our model in a standard encoder-decoder structure. Our feature encoder network consists of two long short-therm memory layers (LSTM)~\cite{hochreiter1997long}, followed by a decoder network, which outputs a probability distribution of classes. We use the deep learning framework PyTorch for implementing the LSTM layers, along with PyTorch-Lightning for module and training cycle management. We formulate the classification task as a regression problem by using ordinal labels~\cite{cheng2008neural}. For further details, we refer the reader to \autoref{appendix:data_prep}. We use the mean-squared error (MSE) as a loss function to regress ordinal labels. We use the following hyperparameters: learning rate=$0.00001$, learning rate decay=$0.99$, batch size=$16$, and dropout=$0.5$. For the two LSTM layers, we use a hidden state size of $64$ and a sequence length of $30$. Also, we use every \nth{10} data point, corresponding to a sequence length of $10$ seconds. In our hyperparameter search (see \autoref{app:hyper}), we found the selection of these two hyperparameters to be the best tradeoff between the length of the input sequence and the oversampling of the same values.



\noindent\textbf{Results.} The training of our network resulted in an accuracy of \seven$=59.3\%$, \three$=83.5\%$, and \two$=71.9\%$, outperforming the RF classifier, see \autoref{tab:performances}, which is an expected result regarding the increased number of parameters of our recurrent classifier and the ability to extract neural representations of sequence data. 

\subsection{Vision-based Thermal Comfort Estimation}
\label{ssec:RCNN_classifier}
In this experiment, we evaluated estimating thermal comfort with \textbf{additional} visual data. 
Hence, an additional model was implemented to also incorporate RGB and body pose key points. 
While training an LSTM with normalized RGB tensors is possible, it introduces high redundancy and complexity as individual pixel values are processed for a sequence of images. In past works, architectures for tasks such as action recognition have been proposed, which include a feature extraction step before the LSTM component~\cite{ullahActionRecognitionVideo2018}. For image feature extraction, architectures that employ CNNs are often used~\cite{osheaIntroductionConvolutionalNeural2015}. To obtain image features from each participant's RGB frames, ResNet~\cite{heDeepResidualLearning2016} was selected due to its ability to filter deep features in images while also introducing mechanisms to avoid the vanishing gradient problem. The PyTorch framework provides pre-trained ResNet versions~\cite{pytorchResNet} that can be incorporated directly into existing models. We train our network with a batch size of $4$ and alter the skip rate from $10$ to $3$ to achieve sufficient training speed for the image-based RNN model. Image feature vector extracted by the ResNet with size $512$ and the ten body pose key points (given by OpenPose~\cite{openpose}) are concatenated to the feature vector. We stick to the best-performing feature combination, described in Section~\ref{ssec:LSTM_classifier}.\\
\noindent\textbf{Results.} The classifier reached accuracies of \three$=48.5\%$, \three$=69.4\%$, \two$=76.0\%$, falling behind our classifier trained without image features. A quantitative comparison between all models is reported in \autoref{tab:performances}.

\subsection{Forecasting using Recurrent Classifier}\label{ssec:forecasting}
In our experiments, we apply recurrent neural networks (RNNs) to sequential data to predict the state of a future point in time for a given sequence of the current time step. As RNNs are also successfully used for time series forecasting~\cite{connor1991recurrent,di2017recurrent}, we investigate different forecasting ranges to explore the accuracy of predictions using various forecasting windows. We adopt the classifier presented in Section~\ref{ssec:LSTM_classifier} and train it using labels from time steps later in the future. The forecasting gaps used in our experiments range from $10$ seconds, $5$ min, and $10$ min into the future. 
We train our classifier using the identical training protocol as it is described in Section~\ref{ssec:LSTM_classifier}.\\
\noindent\textbf{Results.} In \autoref{tab:performances}, performance results for our forecasting experiments are reported in row two until row four. The prediction performance between the $10$-second window and the $5$-min window differed only slightly: \seven$=56.6\%$, \three$=80.4\%$, \two$=74.4\%$ for the $10$-second window, \seven$=55.5\%$, \three$=77.7\%$, \two$=74.6\%$ for the $5$-min window. And the $10$-min shows lower accuracies, \seven$=50.9\%$, \three$=77.1\%$, \two$=73.3\%$, than both previous forecasting performances. Experimental runs with forecasting windows exceeding $10$ min showed further decreases in prediction performance.

\subsection{Feature Combination Study}\label{ssec:feature_combination_study}
To assess the combination of input features to our networks, we conducted a feature combination study based on the top-k ranked features by the feature importance method (see \autoref{fig:impurity_imp}). We selected $k=5$, covering the most important features while also yielding a manageable number of combinations. For the experiment, we selected $n$ features out of $k$, where $n\in[3,4,5]$. In total, this results in $16$ combinations of input features. For each training run, we use the LSTM model with identical hyperparameters, as described in Section~\ref{ssec:LSTM_classifier}. \\
\noindent\textbf{Results.} In \autoref{tab:feature_combination_study_results}, we report performance results of all training runs. These results indicate that the best prediction results can be achieved using the following input features: \textit{Galvanic Skin Response, Relative Humidity, Ambient Temperature, and Wrist Skin Temperature}, as demonstrated in Section~\ref{ssec:LSTM_classifier}. Using the five most important features reduces \seven of the classifier to $55.9\%$, \three to $79.8\%$, and \two to $72.1\%$. For detailed analysis results of the feature combination study, incorporating up to $8$ features, as well as an investigation of SHAP values~\cite{lundberg2017unified}, we refer the reader to \autoref{sec:detailed_feature_combination_study}.

\begin{table}[ht]
\setlength{\tabcolsep}{3.5pt}
\centering
\footnotesize
\caption{In this experiment, we report the accuracy of our classification model, ablating different combinations of input features. We highlight the feature combination performing best. During the feature combination study, we investigate the top $5$ ranked features by our feature importance analysis, \autoref{fig:impurity_imp}: Radiation temperature (\textsc{RT}), wrist skin temperature (\textsc{WS}), galvanic skin response (\textsc{GL}), ambient temperature (\textsc{AT}), relative humidity (\textsc{HU}). In the last row, we include baseline results of a null model outputting the most frequent label only.}\label{tab:feature_combination_study_results}
\begin{tabular}{cccccccc}
\toprule
\multicolumn{5}{c}{Features}& \multicolumn{3}{c}{Classification} \\
GL & AT & HU & RT & WS & \seven & \three & \two \\
\midrule
X & X & X &  & X & \textbf{59.3\%} & \textbf{83.5\%} & 71.9\% \\
 & X & X & X & X & 59.1\% & 78.4\% & 73.5\%  \\
 & X &  & X & X & 57.7\% & 79.1\% & 73.8\%   \\
 & X & X &  & X & 57.1\% & 79.6\% & 72.7\%   \\
X & X &  & X & X & 56.2\% & 79.2\% & 72.8\%  \\
X & X & X & X & X & 55.9\% & 79.8\% & 72.1\% \\
X & X &  &  & X & 55.8\% & 79.4\% & 72.8\%   \\
X & X &  & X &  & 55.4\% & 79.2\% & 72.3\%   \\
X & X & X & X &  & 55.0\% & 78.0\% & 72.7\%  \\
 & X & X & X &  & 52.5\% & 78.6\% & 69.6\%   \\
X & X & X &  &  & 50.5\% & 73.9\% & 73.1\%   \\
X &  &  & X & X & 47.7\% & 70.8\% & \textbf{74.6\%}   \\
 &  & X & X & X & 45.6\% & 70.5\% & 70.8\%   \\
X &  & X & X &  & 42.0\% & 64.9\% & 69.0\%   \\
X &  & X & X & X & 41.5\% & 67.5\% & 70.5\%  \\
X &  & X &  & X & 41.1\% & 67.3\% & 68.1\%   \\ 
\midrule                                                     
\multicolumn{5}{l}{Null Model} & 24.8\% & 24.8\% & 63.2\% \\ 
\bottomrule                                             
\end{tabular}%

\end{table}

\subsection{Experiments On In-Vehicle and Combined Dataset}
\label{ssec:experiments_vehicle}

\begin{table}[ht]
\setlength{\tabcolsep}{3.5pt}
\centering
\footnotesize
\caption{Comparing experimental results of five models trained on different datasets. The upper table shows performance results for the in-vehicle test dataset. The models in the bottom rows are tested using the in-lab test dataset. In this experiment, all models were trained using five input features: galvanic skin response, heart rate, relative humidity, ambient temperature, and skin temperature. Note that the evaluation reported in \autoref{tab:feature_combination_study_results} used a different combination of features (radiation temperature used instead of heart rate, as radiation temperature was not available in-vehicle). Additionally, we include baseline results of a null model outputting the most frequent label only.}\label{tab:in_vehicle_results}
\begin{tabular}{ccccc}
\toprule
Train & Evaluate & \seven & \three & \two \\
\midrule
\indoor  & \vehicle & 39.2\% & 55.0\% & 67.5\% \\
\vehicle & \vehicle & 45.2\% & \textbf{63.8\%} & 67.7\% \\ 
Combined & \vehicle & \textbf{48.1\%} & 61.4\% & \textbf{69.3\%} \\
Null Model & \vehicle & 29.6\% & 42.4\% & 64.0\% \\
\hdashline
\indoor  & \indoor  & 52.0\% & 76.4\% & 71.3\% \\ 
Combined & \indoor  & \textbf{56.9\%} & \textbf{79.7\%} & \textbf{72.6\%} \\
Null Model & \indoor & 24.8\% & 24.8\% & 63.2\% \\
\bottomrule                                             
\end{tabular}%
\end{table}

Having collected a real scenario dataset, as described in Section~\ref{sec:vehicle_dataset}, enables us to investigate thermal comfort estimation under real conditions and examine the gap to indoor conditions. To do so, we train two additional classifiers using the same LSTM model architecture and the same hyperparameters as described in Section~\ref{ssec:feature_combination_study} with the best feature combination. We use data from $16$ participants during training and the remaining data for evaluation. In exception, during the in-vehicle study, five sensory measurements are recorded: galvanic skin response, heart rate, relative humidity, ambient temperature, and skin temperature, which are used as input to our classifier. In this experiment, we compared and combined the scenarios in-vehicle and buildings. Therefore, we evaluate all classifiers on our \indoorDataset originating from the in-vehicle study, as described in Section~\ref{sec:vehicle_dataset}, and on our \indoorDataset, see Section~\ref{sec:indoor_dataset}.

\noindent\textbf{Results.} First, we trained one classifier solely on the \vehicleDataset, leading to a performance of \seven$=45.2\%$, \three$=63.8\%$, and \two$=67.7\%$. When combining both datasets (\indoor and \vehicleDataset), we achieved an accuracy of \seven$=48.1\%$, \three$=61.5\%$, and \two$=69.3\%$ (see third row in \autoref{tab:in_vehicle_results}). Additionally, we evaluate the classifier, which was trained on the \indoorDataset, using the \vehicleDataset to compare the difference between building and vehicle scenario, and we report the results in the first row of \autoref{tab:in_vehicle_results} (i.e., \seven$=39.2\%$, \three$=55.0\%$, and \two$=67.7\%$). Looking at the evaluation results of the two models, one trained and tested on the \indoorDataset, and the other trained and tested on the \vehicleDataset, there is a performance difference. We attribute the lower accuracy for the \vehicleDataset compared to the \indoorDataset (second vs. fourth row in \autoref{tab:in_vehicle_results}) to the less controlled environment and fewer data entries. Further, looking at the classifier that was trained on \indoorDataset, its performance for the vehicle scenario lags behind the performance of the classifier that was trained using the \vehicleDataset. This shows a gap between building and in-vehicle data. Ultimately, when combining both data sets (third and last row in \autoref{tab:in_vehicle_results}), it shows that both classifiers benefit from additional data and the test performance for \indoorDataset and \vehicleDataset increases. We draw the conclusion from these experiments that there is a gap between both datasets and that it can be narrowed by increasing training data, yet a gap between both scenarios remains.

\section{Discussion}\label{sec:discussion}
In the following, we discuss the results of the dataset acquisition, model training, and feature importance analysis, as well as similarities and differences to previous approaches in the field of state recognition.

\subsection{Temporal Data for Thermal Comfort Estimation}
For our RF classifier, an evaluation accuracy of $47.1\%$ was reached on the seven-point thermal sensation scale. RF models were previously used to predict thermal preference (three-point scale) and thermal state (two-point scale), where accuracies between $76\%$ (three-point)~\cite{liuPersonalThermalComfort2019} - $94\%$ (two-point)~\cite{CHAUDHURI2018391} were achieved using physiological signals. The performance of the RF model employed in this work was similar, with a prediction accuracy of $73.8\%$ for the three-point scale and $67.9\%$ for the two-point scale. However, the LSTM-based classifier achieved a prediction accuracy > $59\%$ on the seven-point scale (see \autoref{tab:performances}). We attribute the difference to the inherent mechanism, as the RF models were not processed in sequences but on an individual line-by-line basis. Therefore, changes over time could not be incorporated into the model's estimation. This is supported by the improved accuracy when employing data-downsampling. Previous RF models for thermal comfort estimation~\cite{CHAUDHURI2018391, liuPersonalThermalComfort2019} did not employ data-downsampling, as data was already labeled more sparsely in comparison to the thermal comfort dataset that was recorded at $30$Hz.

\subsection{Visual Features for Thermal Comfort Estimation}
Comparing LSTM and CNN-LSTM, the LSTM model was optimized using ambient temperature, humidity, skin temperature, and galvanic skin response as input features, which were identified as the four most relevant features for thermal sensation prediction, while CNN-LSTM model features were selected based on real-world applicability. The LSTM architecture achieved a higher evaluation accuracy score than the CNN-LSTM. However, radiation temperature and even physiological features such as heart rate or GSR may not be readily available in future buildings. Furthermore, the meaningfulness of the raw extracted visual features via the ResNet block could be increased by, for instance, applying facial expression or motion detection~\cite{zhangFrownBasedThermalComfort2021}, before concatenation with the remaining features, as this allows the model to predict thermal comfort on more filtered feature representations. Consequently, the LSTM model would not be guaranteed to outperform the CNN-LSTM architecture in a real-world scenario. The achieved accuracy on the dataset ($59.3\%$) should thus be seen as an initial dataset benchmark.

\subsection{Data Acquisition for State Recognition Datasets}
The data acquisition in this work differed from previous approaches. Firstly, a controlled (low fidelity) climate chamber was employed over $60$ min. Additionally, instead of questionnaire-based data collection, data were recorded using a data logging application that enabled direct labeling by the participants. The logging differed from previous approaches (e.g., \cite{liuPersonalThermalComfort2019,luThermalComfortBasedPersonalized2019,maoThermalComfortEstimation2021}) as the synchronization of the various sensory inputs, as well as the labeling interface, were centralized in a single application that allows for dense sampling \textit{without post-labeling}. Therefore, we assume that the labeling accuracy and log timing were greatly increased, as participants could report their current state by performing a minimal number of interactions with a numeric keypad.
Due to the personal thermal comfort zone~\cite{CIUHA2016123, ciuhaThermalComfortZone2017}, which is person-dependent, a calibration of the model during usage seems appropriate. This could be done, for example, by directly asking users (e.g., ``Are you currently feeling cold, comfortable, or hot?'') or by treating the manual adjustment of the A/C as input (``want warmer'', ``want cooler'', with the amplitude of change being an indicator of the strength of this desire). However, \citet{QUINTANA2023109685}, with their cohort comfort model approach, showed a potential avenue to reduce required personal data.

\subsection{State Recognition For Interaction Design}
In line with work by \citet{stampf2022towards}, our work helps determine the current user's state in an AV. Our dataset has relevance both for manual and especially for automated driving, which is one of the raised points by \citet{stampf2022towards}. With this dataset, we contribute one aspect to enable developers to create a ``digital twin'' that could enable novel interaction design. We envision numerous interaction possibilities here, as the uncertainty of the prediction and severity of the deducted action have to be considered. Additionally, there are multiple ways to use these states, for example, by directly adjusting values or asking the user whether they want a value adjusted. These interactions and effects will become highly relevant with ever-better recognition.

\subsection{Limitations and Future Work}\label{ch:limits}
A low-fidelity climatic chamber was built using three (two heaters and one cooler) commercial smart-home A/C units. As the minimum temperature during the recordings was $17.1$°C, cold sensations could not successfully be induced for all participants, which led to an unbalanced label distribution in the dataset. Additionally, radiation types alter comfort modeling, therefore, the heater types most likely had an impact on the reported thermal comfort~\cite{hirnimpact}. Moreover, physiological signals measured using the Empatica E4 wristband were found to have a high outlier rate (up to $13.7\%$) for heart rate and GSR measurements. Additionally, while methods for boredom aversion during the recordings were implemented, we assume that the overall level of boredom in participants remained high due to the length of the recordings. This potential boredom, in turn, may have led to unwanted influences on emotion ratings.
Furthermore, the measurements for body fat and body temperature were taken using commercial-grade measuring tools, likely affecting the accuracy of the recorded values. The recorded dataset also contains a bias due to the repeated heating pattern in all data recordings. Future work should address the dataset bias by adding further data recorded using different heating patterns and a more diverse participant population to provide more balanced data for state recognition. Regarding thermal comfort classification, future architectures should focus on processing visual and environmental features that can be collected reliably in future real-world scenarios.
The comparability to other datasets is also limited because the ASHRAE II \cite{ashrae2} dataset contains international data from singular observations, and our dataset is limited to one country \anon{Germany}. Also, the variance in clothing for the participants was low, given the short timeframe we used to gather data ($3$ weeks in total).
Also, while there is an influence of age on thermal comfort perception~\cite{DELFERRARO2015177, GUERGOVA201180}, our dataset is limited to data from mostly younger participants (\m{24.64}, \sd{3.03}). 
To overcome such limitations observed in the current study on indoor thermal comfort and vehicle experiments, several improvements are proposed:
First, diversifying the participant pool is crucial. Expanding the participant count will enhance the statistical significance of the findings, while including individuals from various demographics will provide insights into diverse thermal comfort perceptions. Secondly, the experimental conditions should be broadened. This can be achieved by implementing different rates of thermal changes, for example, by abruptly or very subtly increasing or decreasing the temperature to further study how this influences individual thermal comfort state. At the same time, future studies should focus on exposing participants to an equally distributed temperature spectrum to study a wide variety of thermal comfort scenarios and conduct longitudinal studies to observe the effects of transient thermal phenomena over time. Thirdly, conducting comparative studies across different environments and internationally could illuminate specific thermal comfort challenges and strategies, highlighting the influence of various external factors on thermal perceptions. Fourthly, to mitigate subjective differences in the raters’ individual temperature range in which they feel comfortable~\cite{martins2022systematic}, relative thermal comfort changes should be investigated. To do so, the minimal and maximal temperature values representing the minimum and maximum of the comfort scale need to be retrieved per participant and used to normalize temperature measurements. Lastly, addressing behavioral and psychological factors through targeted studies would illuminate how individual actions and mental states impact thermal comfort, further enriching our understanding of this multifaceted topic.
Implementing these suggestions could significantly refine the study's approach, leading to a more comprehensive understanding of thermal comfort dynamics in both indoor and vehicular environments.

\section{Dataset and Code Availability}\label{sec:availability}


The code is available at \url{https://anonymous.4open.science/r/autotherm-3CFF} (currently anonymized). The data will also be made available online after acceptance at HuggingFace. Requesters of the RGB data will have to submit details of the research purposes and IRB approval to the authors for dataset access.

\section{Conclusion}\label{sec:conclusion}
This work explored machine-learning-based thermal comfort recognition as a relevant target state space for future AV interaction concepts. First, relevant influencing factors on thermal comfort and the affiliated sensory devices were reviewed and filtered based on their level of obtrusiveness and their prevalence in previous thermal comfort studies used in the field of building ergonomics (e.g., \cite{fanger1970thermal, maoThermalComfortEstimation2021, InfluenceOfBodyTemp1999}). Then, a thermal comfort experiment and a data logging application were designed to record a dataset. This dataset was collected in two scenarios: buildings and in-vehicle. This dataset includes a total of $31$ different features (age, gender, weight, height, body fat, body temperature, clothing, sport, meal timing, tiredness, radiation temperature, ambient temperature, wrist skin temperature, GSR, heart rate, relative humidity, RGB frames, ten pose key points, metabolic rate, air velocity, and two emotion features). A thermal comfort study with $N=18$ participants in a self-built climatic chamber and $N=20$ participants in a BMW 3 Series was conducted. 
The datasets were then analyzed regarding feature importance based on correlation coefficients, model-estimated data impurity scores, and model-estimated permutation scores. The four most relevant features for thermal comfort estimation with the recorded data were ambient temperature, galvanic skin response, skin temperature, and relative humidity. Three different classifiers (RF, LSTM, and CNN-LSTM) were implemented and trained using the recorded dataset to assess the feasibility of thermal comfort recognition. A thermal comfort estimation accuracy of $59.3\%$ in the climatic chamber and $45.2\%$ in the vehicle was achieved using the implemented LSTM model and the dataset labels given on a seven-point scale (combined: $48.1\%$). The resulting classifier configuration was used for further experiments employing state forecasting and different label ranges (three-point and two-point scales). The reduction of labeling scales increased prediction performance in all implemented classifiers. The recorded dataset and the data logging application will be made available publicly to encourage further data recording projects for state recognition by providing a template for user-labeled data acquisition. This work provides the first dataset and reference implementation for state-of-the-art thermal comfort estimation. It helps developers and designers to gain a better understanding of their users, as well as helps introduce novel automated responses to detected states.

\begin{acks}
We thank all study participants. This work was conducted within the project 'SEMULIN', funded by the Federal Ministry for Economic Affairs and Climate Action (BMWK).
\end{acks}

\bibliographystyle{ACM-Reference-Format}
\bibliography{manuscript}

\appendix

\section{Cross-validation of the Recurrent Network Classifier}
\label{sec:cross-validation}
In our experiment from Section~\ref{ssec:LSTM_classifier}, we used data from $16$ participants during the training of our models. Then, we evaluate performance on the data from the remaining $2$ participants. Now, we investigate the ability of our network to generalize to unseen data. Therefore, we adopt $9$ fold cross-validation, repeating the optimization of our classifier. To do so, we create $9$ folds with two participants each for evaluation. The remaining participants are used for training. Then, we train $9$ classifiers, using the same hyperparameters overall training runs. In \autoref{tab:cross-validation}, we report individual performance measures, as well as the mean and standard deviation for all runs. The mean accuracy of our classifier is \seven$=54.8\%\pm1.7\%$, \three$=78.2\%\pm1.7\%$ and \two$72.3\%\pm0.7\%$.
\begin{table}[ht]
\setlength{\tabcolsep}{3.5pt}
\centering
\footnotesize
\caption{Cross-validation results over $9$ folds with $16$ participants in the training split and remaining $2$ participants for testing. In the last two rows, we report the mean and std for all metrics.}
\Description{This table lists the results of the 9 cross-validation runs. The first column starting from the left, includes the run ID. The following three columns list the achieved classification accuracies for the 7-point, 3-point, and 2-point scales. The final two columns list the achieved mean-squared-error and L1 metrics. The mean accuracy of our LSTM classifier on the 7-point scale is 55.7\% with a standard deviation of 1.0\%, while a mean accuracy of 79.1\% with a standard deviation of 1.2\% was achieved on the 3-point scale. On the 2-point scale a mean accuracy of 72.3\% and a standard deviation of 0.5\% was achieved.}\label{tab:cross-validation}

\begin{tabular}{cccc}
\toprule
    & \multicolumn{3}{c}{Classification} \\
Run & 7-point Acc & 3-point Acc & 2-point Acc \\
\midrule
$0$   & 54.4\% & 79.4\% & 70.9\% \\
$1$   & 52.3\% & 76.0\% & 72.8\% \\
$2$   & 57.0\% & 79.8\% & 72.8\% \\
$3$   & 54.3\% & 77.7\% & 72.1\% \\
$4$   & 57.8\% & 81.1\% & 72.9\% \\
$5$   & 55.6\% & 79.0\% & 71.9\% \\
$6$   & 54.7\% & 77.9\% & 72.5\% \\
$7$   & 53.8\% & 77.9\% & 71.4\% \\
$8$   & 52.9\% & 75.2\% & 73.2\% \\
\hline
Mean   & 54.8\% & 78.2\% & 72.3\% \\
Std   & 1.7\% & 1.7\% & 0.7\% \\
\bottomrule                                             
\end{tabular}%

\end{table}

\section{Ambient Temperature}

\begin{figure}[ht]
        \centering
        \includegraphics[width=0.5\textwidth]{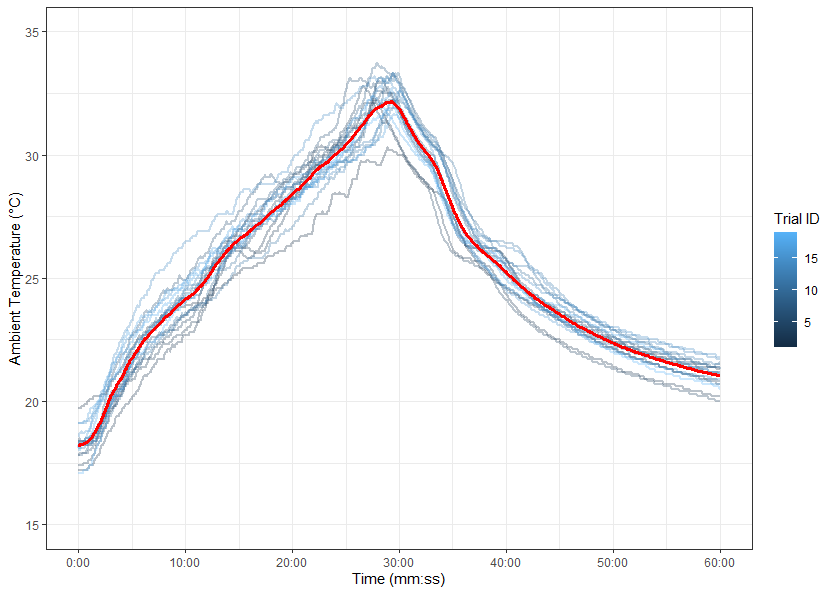}
        \caption{\label{fig:heating_curves} Ambient temperature conditions for the full recording duration during the \indoor study, grouped by trial ID. The mean heating profile is included (red).}
        \Description{This Figure shows a plot that visualizes the ambient temperature profiles that were achieved over for each participant during the thermal comfort study. The plot shows that a minimum temperature of 17.1°C and a maximum temperature of 33.7°C was reached. The mean temperature profile across all participants is shown in red.}
\end{figure}

\autoref{fig:heating_curves} illustrates the ambient temperature profile that each participant was exposed to. It can be observed that the defined temperature ranges were met in each trial during the recording sessions.

\section{Data Preparation and Augmentation}\label{appendix:data_prep}
Firstly, outliers were defined as outside of the mean with three times the standard deviation. This was also done per label group for the ambient temperature. That means that for a given label, data points with an ambient temperature of $+/- 3*std$ of the mean were excluded. This was done to filter falsely attributed labels. Then, we used one-hot encoding for categorical features and overall data down-sampling for the RF classifier.

For the LSTM and CNN-LSTM models, the same outlier filtering, one-hot-encoding, and down-sampling steps were employed. However, as sequences are expected as input to the LSTM models, a sliding window approach was implemented, where, for every index in the created data frame, a sequence of length $n$ is created, such that, for instance, a window size of $w=30$ would result in the following data frame scheme.

Simple data augmentation for continuous features was also implemented. Gaussian noise sampled from a Gaussian distribution with the parameters $\mu=0.00$ and $\delta=0.30$ was added to the continuous variables through element-wise addition. For image data, a more extensive data augmentation scheme was applied, as RGB frames were recorded at a size of 1920x1080, which is too large for efficient training. Thus, first, a central 1000x1000 crop was extracted and resized to 224x224. The resized image is then randomly rotated up to 5°, randomly flipped horizontally, and then normalized so that all three RGB channels are given as values between 0 and 1

Finally, as the thermal sensation scale has an ordinal scaling level, providing simple integer targets as prevalent in the raw dataset ignores the rank information. Therefore, the labels were changed according to a scheme proposed by \citet{cheng2008neural}. Ordinal labels are transformed into binary vectors of size $k$ (here 7), where $k$ is the number of ranks given on the ordinal scale. The binary vectors are instantiated as zero-vectors and then filled with ones from left to right, depending on the rank of the ordinal label. The resulting label transformations for the thermal sensation scale (seven-point scale) were: 

\begin{itemize}[noitemsep]
    \item $-3 \longrightarrow [1,0,0,0,0,0,0]$ 
    \item $-2 \longrightarrow [1,1,0,0,0,0,0]$ 
    \item $-1 \longrightarrow [1,1,1,0,0,0,0]$ 
    \item $ 0 \longrightarrow [1,1,1,1,0,0,0]$
    \item $ 1 \longrightarrow [1,1,1,1,1,0,0]$ 
    \item $ 2 \longrightarrow [1,1,1,1,1,1,0]$ 
    \item $ 3 \longrightarrow [1,1,1,1,1,1,1]$
\end{itemize} 

The resulting binary vectors were used as labels for the LSTM and CNN-LSTM models. Split sizes were predefined, so data from 16 participants were used for training, and the remaining data files for validation/testing. 

\begin{figure*}[ht]
  \begin{subfigure}{0.45\textwidth}
    \centering
    \includegraphics[width=\linewidth]{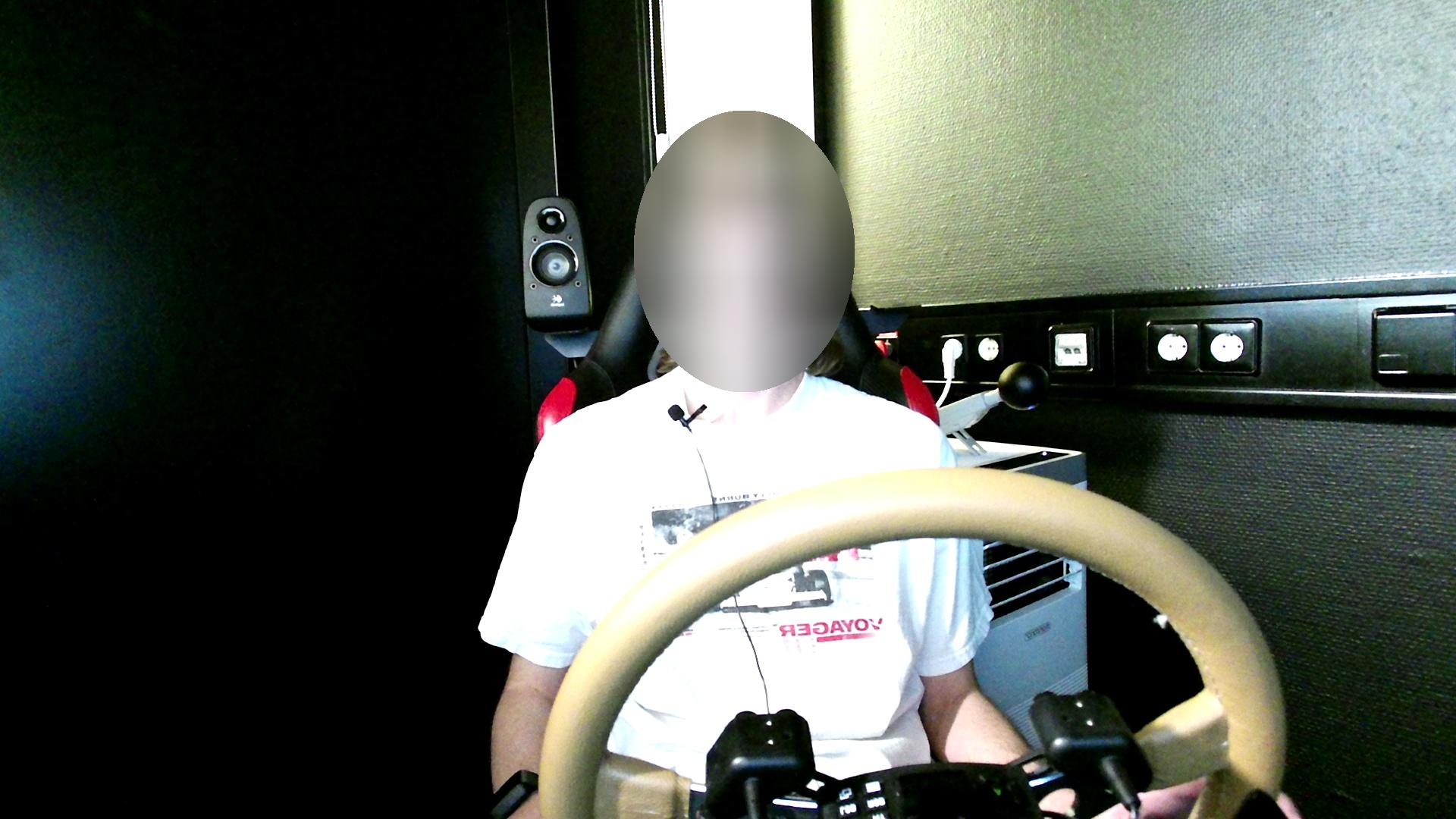}
    \caption*{In-Lab study}
  \end{subfigure}%
  \begin{subfigure}{0.45\textwidth}
    \centering
    \includegraphics[width=\linewidth]{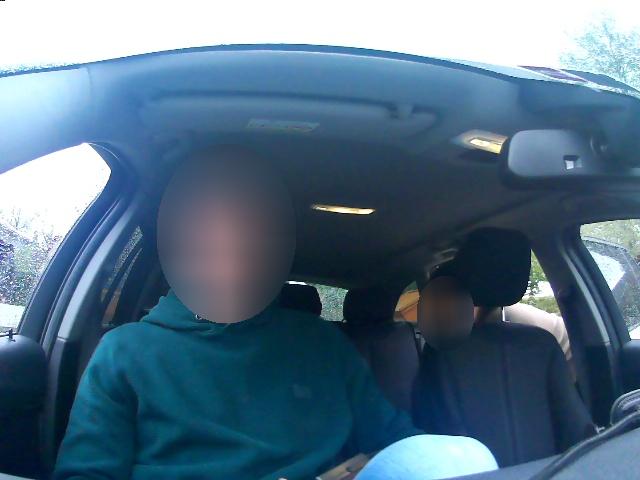}
    \caption*{In-Vehicle study}
  \end{subfigure}
 
  \caption{Left: Image taken from the setup during the in-lab study. Right: Image taken from inside the BMW 3 during the in-vehicle study.}
  \label{fig:person-in-lab-in-vehicle}
  \Description{}
\end{figure*}

\section{LSTM Hyperparameter Search}\label{app:hyper}
For the LSTM model (without visual features), initial trial runs on the data showed severe over-fitting symptoms (i.e., training loss steadily declining, while validation loss increased continuously). The model was then modified such that the hidden state size of the LSTM cells and the input sequence length could be included in the hyperparameter search. The optimization parameters for the initial LSTM model, therefore, included batch size, learning rate, sequence length, down-sampling rate, drop-out, number of LSTM layers, and the hidden state size. A training script using the parameter ranges shown in \autoref{tab:lstm_search} was employed, and each configuration was run initially run for 20 epochs. The mean-squared error (MSE) was used as a loss function. The highest validation accuracy using the fore-named parameters was achieved at 0.62 with the configuration: learning rate=0.00001, hidden states=64, sequence length=300 (i.e., 10 seconds), batch size=16, layers=2, and drop-out=0.5. The same feature combination was used for all the runs. After analyzing the loss graphs for the training and validation data, it became notable that improvement of the model decreases quickly after the learning rate decreases far below 0.00001. Moreover, over-fitting was less severe with smaller hidden state sizes ([32-64]). One reason for the quick drop in model improvement was assumed to be the learning rate decay, which was thus far implemented as a reduction by $x^{0.99999}$. To explore the learning rate decay's influence on the model improvement, it was also added to the optimization parameters.

\begin{table}[ht]
\scriptsize
\def\arraystretch{2}
\centering
\begin{tabular}{|p{3cm}|p{8cm}|}
\hline
\textbf{Parameter} & \textbf{Range}\\ \hline
batch size & \{4, 8, 16\}\\ \hline
learning rate & \{0.001, 0.0001, 0.00005, 0.00001\}\\ \hline
learning rate decay & \{0.99999, 0.999999, 0.9999999\}\\ \hline
hidden states& \{10, 16, 32, 64, 96, 128, 256\}\\ \hline
lstm layers& \{1, 2, 3\}\\ \hline
sequence length& \{10, 50, 100, 150, 300, 600\}\\ \hline
drop-out&\{0.2, 0.5\}\\ \hline
features&\{clothing, radiation temperature, ambient temperature, ambient humidity, GSR, heart rate, skin temperature\}\\ \hline
\end{tabular}
\caption{\label{tab:lstm_search} The parameter space used for grid search optimization of the LSTM and CNN-LSTM models.}
\end{table}

Keeping the previously optimized parameters the same while varying only learning rate decay and down-sampling rate, additional training runs were started and trained for up to a maximum of 100 epochs or until over-fitting was detected. The number of training epochs was increased as a response to the low learning rate and decreased learning rate decay. Optimization of the remaining parameters resulted in a validation accuracy of 0.71 with the configuration learning rate=0.00001, hidden states=64, layers=2, down-sampling=10, sequence length=30 (i.e., 10 seconds), batch size=16, and drop-out=0.5, and learning rate decay=0.9999999. 

\section{Feature Permutation Importance}\label{sec:feature-permutation-importance}

\begin{figure*}[ht]
  \begin{subfigure}{\textwidth}
    \centering
    \includegraphics[width=\textwidth]{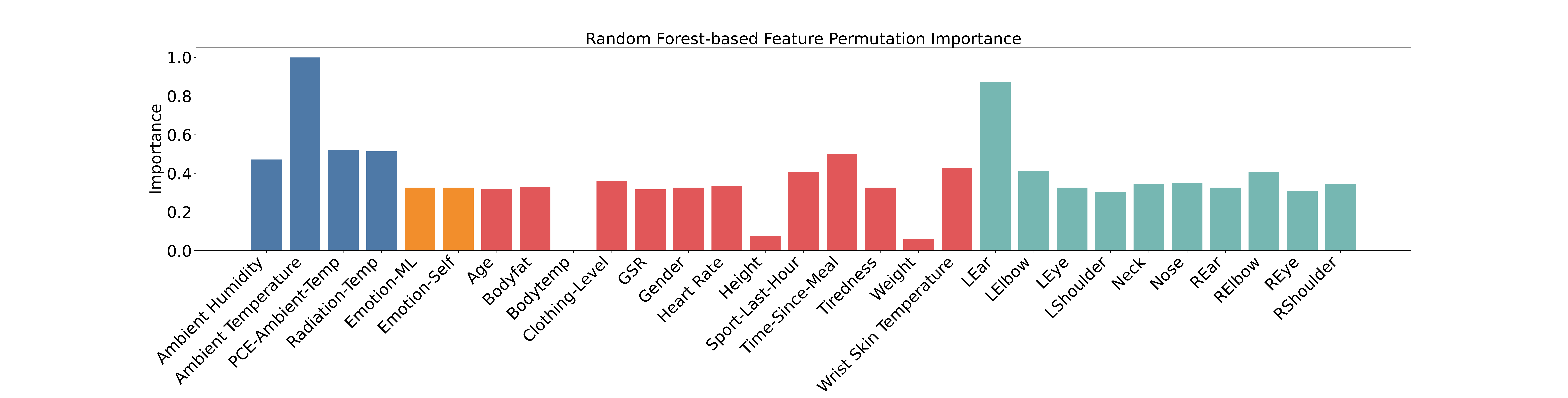}
  \end{subfigure}%
  \\
  \begin{subfigure}{\textwidth}
    \centering
    \includegraphics[width=\textwidth]{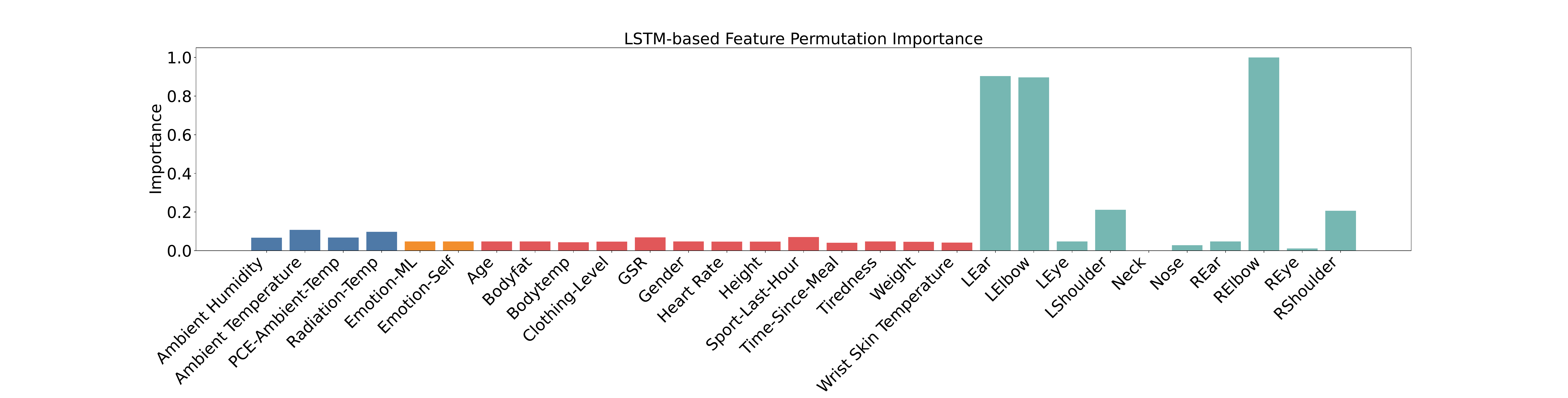}
  \end{subfigure}
 
  \caption{Feature permutation importance is computed for both models, (top) Random Classifier and (bottom) LSTM-base model. We group by \textcolor{bluegray}{physical}, \textcolor{darkorange}{psychological}, \textcolor{darkpastelred}{physiological} and \textcolor{etonblue}{OpenPose}~\cite{openpose} features. While the results for the random forest indicate more importance to the physical and some physiological features, for the LSTM-based similar conclusion can be drawn, if only the differences are only marginal. However, features derived from RGB Frames using OpenPose~\cite{openpose} seem to matter only in some cases (left ear for random forest and left ear, and elbows for the LSTM model). We attribute this inconsistency to occluded body parts in the recordings due to sitting posture.}
  \label{fig:feature-permutation-importance}
  \Description{}
\end{figure*}

\newpage

\section{Extended Feature Combination Study}\label{sec:detailed_feature_combination_study}
\begin{figure*}[ht]
        \centering
        \includegraphics[width=\textwidth]{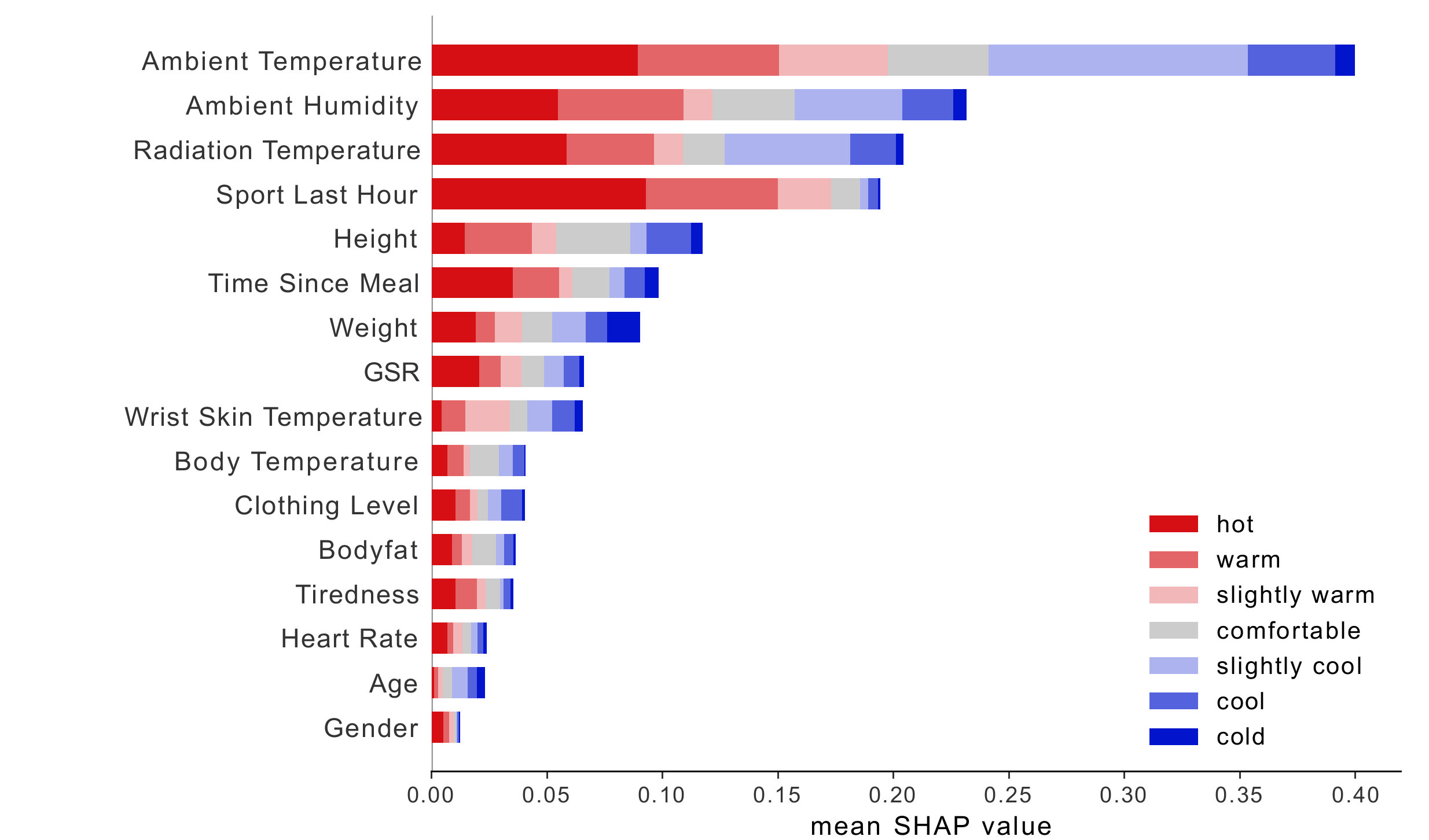}
        \caption{\label{fig:shap_values} SHAP values for sixteen features computed for the test split of the \indoorDataset using the trained RF. For each feature and class, we report the mean absolute SHAP value.}
        \Description{SHAP values for sixteen features computed for the test split of the \indoorDataset.}
\end{figure*}

In this experiment, we investigate the importance of different features. We train $255$ classifiers using all possible feature combinations using the following $8$ features: Body temperature (BT),  PCE ambient temperature (PCE), heart rate (HR),  galvanic skin response (GL), ambient temperature from Arduino sensor (AT), relative humidity from Arduino sensor (HU), radiation temperature (RT), wrist skin temperature (WS). We report performance results for classification and regression measures in \autoref{tab:feature_combination_study_large_results_top_50}, \autoref{tab:feature_combination_study_large_results_51_to_100}, \autoref{tab:feature_combination_study_large_results_101_to_150}, \autoref{tab:feature_combination_study_large_results_151_to_200}, \autoref{tab:feature_combination_study_large_results_201_to_250} and \autoref{tab:feature_combination_study_large_results_251_to_255}. Additionally, we compute the absolute SHAP values~\cite{lundberg2017unified}, averaged for the \indoorDataset, see \autoref{fig:shap_values}.
\begin{table}[ht]
\setlength{\tabcolsep}{3.5pt}
\centering
\scriptsize
\caption{Top 50 runs of feature combination study.}\label{tab:feature_combination_study_large_results_top_50}
\begin{tabular}{ccccccccccc}
\toprule
\multicolumn{8}{c}{Features}& \multicolumn{3}{c}{Classification}  \\
BT & PCE & HR & GL & AT & HU & RT & WS & 7-point Acc & 3-point Acc & 2-point Acc \\
\midrule
 &  &  & X & X & X &  & X & 59.3\% & 83.5\% & 71.9\%        \\
 &  &  &  & X & X & X & X & 59.1\% & 78.4\% & 73.5\%        \\
 &  &  &  & X &  & X & X & 57.7\% & 79.1\% & 73.8\%         \\
 & X &  &  & X &  & X & X & 57.5\% & 78.4\% & 74.4\%        \\
 &  &  &  & X & X &  & X & 57.1\% & 79.6\% & 72.7\%         \\
 & X &  & X & X & X &  & X & 57.2\% & 80.3\% & 72.4\%       \\
 & X &  & X & X & X & X & X & 56.5\% & 80.3\% & 72.0\%      \\
 & X &  &  & X &  &  & X & 56.5\% & 79.4\% & 73.3\%         \\
 & X &  & X & X &  & X & X & 56.4\% & 81.5\% & 72.4\%       \\
 &  &  & X & X &  & X & X & 56.2\% & 79.2\% & 72.8\%        \\
 & X &  &  & X & X &  & X & 56.1\% & 79.2\% & 72.9\%        \\
 & X & X & X & X & X &  & X & 56.0\% & 79.5\% & 73.2\%      \\
 &  &  & X & X & X & X & X & 55.9\% & 79.8\% & 72.1\%       \\
 &  &  & X & X &  &  & X & 55.8\% & 79.4\% & 72.8\%         \\
 & X &  & X & X &  & X &  & 55.5\% & 78.8\% & 72.7\%        \\
 &  & X & X & X &  &  & X & 55.5\% & 78.1\% & 73.3\%        \\
 &  &  &  & X &  &  & X & 55.5\% & 79.1\% & 70.9\%          \\
 &  &  & X & X &  & X &  & 55.4\% & 79.2\% & 72.3\%         \\
 & X & X & X & X &  &  & X & 55.3\% & 77.9\% & 73.4\%       \\
 & X &  & X & X &  &  & X & 55.2\% & 78.3\% & 72.6\%        \\
 & X & X &  & X & X &  & X & 54.8\% & 78.5\% & 72.4\%       \\
 &  &  &  & X &  & X &  & 54.5\% & 78.8\% & 72.1\%          \\
 &  & X & X & X &  & X & X & 54.4\% & 77.5\% & 72.9\%       \\
 & X &  &  & X & X & X & X & 54.4\% & 78.9\% & 71.0\%       \\
 &  & X &  & X &  &  & X & 54.3\% & 77.6\% & 72.0\%         \\
 & X &  &  & X &  &  &  & 54.3\% & 78.8\% & 69.9\%          \\
 & X &  & X & X &  &  &  & 54.1\% & 78.5\% & 71.6\%         \\
 & X &  & X & X & X & X &  & 54.1\% & 78.9\% & 71.7\%       \\
 & X &  &  & X & X & X &  & 53.9\% & 77.5\% & 71.7\%        \\
 &  & X & X & X &  & X &  & 53.8\% & 76.9\% & 72.3\%        \\
 &  & X &  &  & X & X & X & 53.8\% & 77.7\% & 71.2\%        \\
 &  & X & X & X & X & X & X & 53.6\% & 77.0\% & 72.6\%      \\
 & X & X &  & X &  &  & X & 53.3\% & 76.9\% & 72.5\%        \\
 & X &  &  & X & X &  &  & 53.3\% & 76.9\% & 71.7\%         \\
 &  & X &  & X & X &  & X & 52.9\% & 77.3\% & 70.7\%        \\
X & X &  & X & X & X &  & X & 52.8\% & 76.4\% & 72.2\%      \\
 & X & X &  & X &  & X &  & 52.6\% & 75.6\% & 72.3\%        \\
 & X &  &  & X &  & X &  & 52.6\% & 76.6\% & 73.5\%         \\
 &  &  &  & X & X &  &  & 52.4\% & 75.4\% & 71.3\%          \\
 & X & X & X & X &  & X &  & 52.3\% & 76.8\% & 70.9\%       \\
 &  &  & X & X &  &  &  & 52.1\% & 74.4\% & 73.5\%          \\
 & X &  & X & X & X &  &  & 52.1\% & 80.5\% & 68.3\%        \\
 &  & X &  & X &  & X & X & 52.1\% & 76.5\% & 70.6\%        \\
 &  &  &  & X &  &  &  & 52.1\% & 74.7\% & 72.6\%           \\
 &  & X & X & X & X &  & X & 52.0\% & 76.4\% & 71.3\%       \\
 & X & X & X & X &  & X & X & 51.9\% & 75.9\% & 72.4\%      \\
 & X &  & X &  & X &  & X & 51.7\% & 76.3\% & 71.3\%        \\
 &  & X &  & X & X & X & X & 51.3\% & 77.2\% & 68.9\%       \\
 &  & X &  & X &  & X &  & 51.3\% & 74.1\% & 72.6\%         \\
 &  &  &  & X & X & X &  & 51.3\% & 78.2\% & 68.3\%         \\

\bottomrule                                             
\end{tabular}%

\end{table}

\begin{table}[ht]
\setlength{\tabcolsep}{3.5pt}
\centering
\scriptsize
\caption{Top 51 - 100 runs of feature combination study.}\label{tab:feature_combination_study_large_results_51_to_100}
\begin{tabular}{ccccccccccc}
\toprule
\multicolumn{8}{c}{Features}& \multicolumn{3}{c}{Classification} \\
BT & PCE & HR & GL & AT & HU & RT & WS & 7-point Acc & 3-point Acc & 2-point Acc \\
\midrule

 &  & X &  & X &  &  &  & 51.3\% & 74.0\% & 72.7\%          \\
X & X & X &  &  & X & X & X & 51.2\% & 78.1\% & 69.5\%      \\
X &  & X &  & X &  &  &  & 51.2\% & 74.6\% & 72.3\%         \\
 &  & X &  & X & X &  &  & 51.2\% & 75.2\% & 71.6\%         \\
 & X & X & X &  & X & X & X & 51.0\% & 74.3\% & 71.4\%      \\
X &  & X &  & X & X & X & X & 50.9\% & 77.2\% & 69.9\%      \\
X & X & X & X & X &  &  & X & 50.8\% & 76.8\% & 70.6\%      \\
 &  &  & X & X & X &  &  & 50.8\% & 75.3\% & 72.1\%         \\
 & X & X &  & X &  & X & X & 50.7\% & 76.4\% & 69.8\%       \\
X & X & X & X & X & X &  & X & 50.7\% & 77.4\% & 69.9\%     \\
 & X & X & X & X & X &  &  & 50.6\% & 73.7\% & 73.0\%       \\
 &  & X & X &  &  & X & X & 50.6\% & 73.9\% & 72.3\%        \\
 & X & X & X &  &  &  & X & 50.6\% & 75.2\% & 70.9\%        \\
 & X & X &  & X &  &  &  & 50.5\% & 76.2\% & 71.8\%         \\
X & X & X & X &  & X & X & X & 50.4\% & 74.7\% & 71.5\%     \\
 & X & X & X & X &  &  &  & 50.3\% & 72.8\% & 73.1\%        \\
X & X & X & X &  & X &  & X & 50.3\% & 75.0\% & 71.3\%      \\
X &  & X & X &  & X & X & X & 50.2\% & 75.0\% & 71.1\%      \\
 &  &  & X & X & X & X &  & 55.0\% & 78.0\% & 72.7\%        \\
X & X & X &  & X & X &  & X & 50.0\% & 75.6\% & 70.8\%      \\
 & X & X & X &  &  & X & X & 49.8\% & 72.8\% & 73.2\%       \\
 & X & X &  & X & X & X & X & 49.8\% & 73.9\% & 70.7\%      \\
 & X & X &  &  & X & X & X & 49.8\% & 73.2\% & 71.7\%       \\
 & X & X & X & X & X & X & X & 49.8\% & 74.9\% & 70.2\%     \\
 & X & X &  & X & X & X &  & 49.6\% & 72.2\% & 73.1\%       \\
X &  & X &  &  &  & X & X & 49.5\% & 72.1\% & 71.7\%        \\
 & X & X &  &  & X &  & X & 49.4\% & 73.5\% & 70.8\%        \\
 & X & X &  &  &  &  & X & 49.3\% & 72.8\% & 73.7\%         \\
 & X &  & X &  &  & X & X & 49.0\% & 71.6\% & 73.6\%        \\
 &  & X & X & X &  &  &  & 48.7\% & 71.5\% & 72.9\%         \\
 &  & X &  & X & X & X &  & 48.7\% & 78.5\% & 69.5\%        \\
 &  &  & X &  &  & X & X & 48.5\% & 70.8\% & 70.2\%         \\
 &  & X &  &  &  & X & X & 48.5\% & 71.5\% & 72.5\%         \\
X &  &  & X &  & X & X & X & 48.2\% & 70.5\% & 73.8\%       \\
X & X & X & X & X & X & X & X & 48.2\% & 73.4\% & 69.2\%    \\
 & X & X &  &  &  & X & X & 48.2\% & 70.8\% & 72.2\%        \\
X & X & X &  &  & X &  & X & 48.2\% & 71.7\% & 71.4\%       \\
 & X &  & X &  &  &  & X & 48.1\% & 72.0\% & 73.3\%         \\
 & X &  & X &  & X & X & X & 48.0\% & 73.7\% & 70.8\%       \\
 &  & X & X &  & X & X & X & 48.0\% & 71.7\% & 72.4\%       \\
X & X & X &  & X &  &  & X & 48.0\% & 75.1\% & 68.6\%       \\
X & X &  &  &  & X &  & X & 47.9\% & 73.8\% & 69.0\%        \\
X &  &  &  &  & X &  & X & 47.9\% & 74.7\% & 68.5\%         \\
 & X &  &  &  &  &  & X & 47.7\% & 70.6\% & 73.1\%          \\
X &  & X & X & X & X & X & X & 47.7\% & 73.3\% & 68.9\%     \\
X & X & X &  & X &  & X & X & 47.7\% & 71.9\% & 71.2\%      \\
 &  & X & X & X & X &  &  & 47.6\% & 74.7\% & 72.2\%        \\
X & X &  &  & X & X &  & X & 47.6\% & 72.6\% & 66.9\%       \\
 & X &  &  &  & X & X & X & 47.5\% & 71.6\% & 71.7\%        \\
X & X &  & X &  & X &  & X & 47.4\% & 72.5\% & 70.7\%       \\

\bottomrule                                             
\end{tabular}%

\end{table}

\begin{table}[ht]
\setlength{\tabcolsep}{3.5pt}
\centering
\scriptsize
\caption{Top 101 - 150 runs of feature combination study.}\label{tab:feature_combination_study_large_results_101_to_150}
\begin{tabular}{ccccccccccc}
\toprule
\multicolumn{8}{c}{Features}& \multicolumn{3}{c}{Classification}  \\
BT & PCE & HR & GL & AT & HU & RT & WS & 7-point Acc & 3-point Acc & 2-point Acc\\
\midrule

X &  & X &  & X &  & X & X & 47.3\% & 73.0\% & 70.4\%       \\
X &  &  &  & X & X &  & X & 47.1\% & 71.2\% & 69.4\%        \\
 &  & X & X & X & X & X &  & 47.1\% & 78.1\% & 68.7\%       \\
X & X & X &  &  &  & X & X & 47.1\% & 72.1\% & 68.9\%       \\
X &  &  &  & X & X & X & X & 46.7\% & 71.0\% & 69.9\%       \\
X &  &  & X &  &  & X & X & 46.6\% & 69.2\% & 73.0\%        \\
X &  & X &  &  & X & X & X & 46.6\% & 70.1\% & 69.9\%       \\
X & X &  &  &  &  & X & X & 46.5\% & 67.4\% & 70.1\%        \\
X &  & X &  & X & X &  & X & 46.4\% & 73.9\% & 68.1\%       \\
X & X &  & X &  &  & X & X & 46.3\% & 67.3\% & 71.0\%       \\
 &  &  &  &  &  & X & X & 46.3\% & 68.8\% & 73.8\%          \\
X & X & X & X &  &  &  & X & 46.3\% & 69.4\% & 71.1\%       \\
X & X &  & X &  &  &  & X & 45.9\% & 67.3\% & 70.0\%        \\
 & X &  &  &  &  & X &  & 45.7\% & 69.5\% & 74.9\%          \\
X &  & X &  & X &  & X &  & 45.7\% & 69.0\% & 73.0\%        \\
 & X & X & X & X & X & X &  & 45.7\% & 77.0\% & 67.9\%      \\
 & X & X &  & X & X &  &  & 45.7\% & 68.5\% & 72.9\%        \\
 & X &  &  &  &  & X & X & 45.7\% & 69.0\% & 73.6\%         \\
X & X &  &  & X &  & X & X & 45.4\% & 70.8\% & 69.7\%       \\
X & X & X &  &  &  &  & X & 45.3\% & 69.7\% & 69.1\%        \\
X &  &  &  &  &  & X & X & 45.3\% & 66.0\% & 72.1\%         \\
X & X &  &  &  & X & X & X & 45.2\% & 69.4\% & 69.0\%       \\
X &  & X & X & X &  &  & X & 45.2\% & 71.4\% & 68.4\%       \\
X &  & X & X &  &  & X & X & 45.0\% & 69.4\% & 71.2\%       \\
 & X & X & X &  & X &  & X & 44.9\% & 69.8\% & 70.8\%       \\
X & X & X &  & X & X & X & X & 44.7\% & 71.1\% & 68.8\%     \\
X & X & X &  & X &  &  &  & 44.4\% & 68.3\% & 71.8\%        \\
X &  &  & X &  & X &  &  & 44.3\% & 71.2\% & 68.9\%         \\
X & X & X & X &  &  & X & X & 44.2\% & 69.5\% & 68.8\%      \\
X &  & X &  &  & X &  & X & 44.2\% & 73.6\% & 68.1\%        \\
X & X &  & X &  & X & X & X & 44.2\% & 69.6\% & 70.7\%      \\
 & X &  & X &  &  & X &  & 44.2\% & 70.1\% & 71.9\%         \\
 & X &  & X &  & X & X &  & 44.1\% & 70.3\% & 66.1\%        \\
X &  & X & X &  & X &  & X & 44.1\% & 62.1\% & 69.0\%       \\
X & X & X & X & X &  & X & X & 43.6\% & 68.7\% & 68.7\%     \\
 & X &  &  &  & X &  & X & 43.6\% & 69.8\% & 69.5\%         \\
X &  &  & X & X &  & X & X & 43.5\% & 66.8\% & 69.9\%       \\
X &  & X & X & X & X & X &  & 43.3\% & 67.8\% & 70.5\%      \\
X &  &  & X & X & X & X & X & 42.8\% & 66.8\% & 71.0\%      \\
X & X & X &  & X & X &  &  & 42.3\% & 67.2\% & 71.6\%       \\
 &  &  & X &  & X & X & X & 42.1\% & 68.5\% & 70.4\%        \\
X &  & X & X & X &  & X & X & 42.0\% & 75.6\% & 62.9\%      \\
X &  &  &  &  & X & X & X & 42.0\% & 70.7\% & 66.9\%        \\
X & X &  & X & X & X &  &  & 41.9\% & 66.6\% & 70.3\%       \\
X &  &  &  & X &  & X & X & 41.9\% & 65.7\% & 69.5\%        \\
X &  &  &  & X &  & X &  & 41.9\% & 66.1\% & 71.5\%         \\
X & X &  &  &  & X &  &  & 41.7\% & 65.8\% & 62.8\%         \\
 &  &  &  &  & X & X &  & 41.6\% & 63.2\% & 71.1\%          \\
X & X &  &  & X &  & X &  & 41.6\% & 66.6\% & 71.5\%        \\
X &  & X &  & X &  &  & X & 41.5\% & 70.0\% & 68.8\%        \\

\bottomrule                                             
\end{tabular}%

\end{table}

\begin{table}[ht]
\setlength{\tabcolsep}{3.5pt}
\centering
\scriptsize
\caption{Top 151 - 200 runs of feature combination study.}\label{tab:feature_combination_study_large_results_151_to_200}
\begin{tabular}{ccccccccccc}
\toprule
\multicolumn{8}{c}{Features}& \multicolumn{3}{c}{Classification}  \\
BT & PCE & HR & GL & AT & HU & RT & WS & 7-point Acc & 3-point Acc & 2-point Acc \\
\midrule

 &  &  &  &  & X & X & X & 41.4\% & 67.3\% & 70.1\%         \\
X &  &  &  & X &  &  & X & 41.4\% & 65.6\% & 66.8\%         \\
 & X & X & X &  & X & X &  & 41.3\% & 69.3\% & 69.1\%       \\
X & X &  &  & X & X & X & X & 41.3\% & 72.6\% & 62.4\%      \\
X &  &  & X &  & X &  & X & 41.3\% & 64.9\% & 61.0\%        \\
X &  &  &  &  & X &  &  & 41.2\% & 64.1\% & 72.2\%          \\
X &  & X & X & X & X &  & X & 40.9\% & 65.0\% & 65.7\%      \\
X & X & X & X & X & X & X &  & 40.8\% & 65.8\% & 71.2\%     \\
 & X & X &  &  & X & X &  & 40.8\% & 68.1\% & 70.5\%        \\
X & X &  &  & X & X &  &  & 40.7\% & 67.2\% & 69.5\%        \\
X &  & X & X & X & X &  &  & 40.7\% & 65.9\% & 70.8\%       \\
X &  & X &  & X & X & X &  & 40.6\% & 66.1\% & 70.8\%       \\
X &  &  & X & X & X &  &  & 40.4\% & 67.6\% & 69.2\%        \\
X &  & X &  & X & X &  &  & 40.4\% & 68.0\% & 68.1\%        \\
 &  &  & X &  & X &  & X & 40.4\% & 67.6\% & 64.7\%         \\
X &  & X & X &  & X &  &  & 40.3\% & 64.0\% & 61.2\%        \\
X & X &  &  & X &  &  & X & 40.2\% & 61.7\% & 61.2\%        \\
X & X & X & X & X & X &  &  & 40.1\% & 63.7\% & 71.1\%      \\
X &  & X & X & X &  & X &  & 40.1\% & 62.9\% & 68.6\%       \\
X & X &  & X & X &  &  & X & 40.1\% & 63.5\% & 68.6\%       \\
X &  &  & X & X &  &  &  & 40.0\% & 65.0\% & 69.2\%         \\
X &  &  &  &  & X & X &  & 39.8\% & 63.0\% & 71.9\%         \\
X & X &  & X & X &  &  &  & 39.7\% & 64.7\% & 69.5\%        \\
X & X &  & X & X & X & X &  & 39.7\% & 64.6\% & 68.5\%      \\
 &  & X &  &  & X &  & X & 39.5\% & 64.7\% & 66.0\%         \\
 &  &  &  &  & X &  & X & 39.4\% & 64.7\% & 63.4\%          \\
X & X & X &  & X &  & X &  & 39.4\% & 64.6\% & 68.8\%       \\
X &  &  & X & X &  &  & X & 39.3\% & 74.1\% & 63.2\%        \\
X &  & X &  &  & X & X &  & 39.3\% & 66.4\% & 64.5\%        \\
X & X & X &  &  & X & X &  & 39.3\% & 65.0\% & 63.7\%       \\
X & X &  & X & X &  & X &  & 39.3\% & 64.3\% & 69.5\%       \\
X &  &  &  & X & X & X &  & 39.0\% & 63.5\% & 70.4\%        \\
X &  & X &  &  & X &  &  & 39.0\% & 65.0\% & 60.3\%         \\
X &  & X & X & X &  &  &  & 39.0\% & 65.6\% & 69.7\%        \\
 & X & X &  &  &  & X &  & 38.7\% & 67.9\% & 72.3\%         \\
X & X &  &  & X &  &  &  & 38.6\% & 63.1\% & 69.3\%         \\
 & X &  &  &  & X & X &  & 38.6\% & 67.9\% & 67.4\%         \\
 &  &  & X &  &  & X &  & 38.5\% & 56.4\% & 73.3\%          \\
X &  &  &  & X & X &  &  & 38.5\% & 63.8\% & 69.0\%         \\
 &  &  & X &  & X & X &  & 38.5\% & 62.2\% & 71.8\%         \\
X &  &  &  & X &  &  &  & 38.5\% & 62.5\% & 69.1\%          \\
X & X & X & X & X &  &  &  & 38.4\% & 65.5\% & 68.9\%       \\
 &  & X &  &  & X & X &  & 38.2\% & 58.2\% & 72.9\%         \\
X &  &  & X & X & X &  & X & 38.1\% & 63.2\% & 63.6\%       \\
X & X &  &  &  & X & X &  & 38.1\% & 60.8\% & 64.9\%        \\
X &  &  & X & X & X & X &  & 38.0\% & 62.7\% & 67.8\%       \\
 & X & X & X &  & X &  &  & 38.0\% & 60.7\% & 70.3\%        \\
 & X & X &  &  & X &  &  & 37.8\% & 59.6\% & 68.5\%         \\
 &  & X & X &  & X & X &  & 37.8\% & 58.9\% & 70.5\%        \\
X & X & X &  &  & X &  &  & 37.8\% & 55.7\% & 66.5\%        \\

\bottomrule                                             
\end{tabular}%

\end{table}

\begin{table}[ht]
\setlength{\tabcolsep}{3.5pt}
\centering
\scriptsize
\caption{Top 201 - 250 runs of feature combination study.}\label{tab:feature_combination_study_large_results_201_to_250}
\begin{tabular}{ccccccccccc}
\toprule
\multicolumn{8}{c}{Features}& \multicolumn{3}{c}{Classification}  \\
BT & PCE & HR & GL & AT & HU & RT & WS & 7-point Acc & 3-point Acc & 2-point Acc\\
\midrule

X &  &  & X & X &  & X &  & 37.7\% & 61.2\% & 69.7\%    \\
 &  &  & X &  & X &  &  & 37.6\% & 65.7\% & 67.5\%      \\
X & X &  &  &  &  &  & X & 37.5\% & 58.9\% & 70.1\%     \\
X & X & X & X &  & X & X &  & 37.4\% & 61.6\% & 64.0\%  \\
X & X &  & X & X & X & X & X & 37.4\% & 61.5\% & 67.4\% \\
X &  & X & X &  & X & X &  & 37.3\% & 56.2\% & 71.0\%   \\
 & X &  & X &  & X &  &  & 37.3\% & 59.4\% & 68.4\%     \\
 &  &  &  &  &  & X &  & 37.1\% & 56.4\% & 74.6\%       \\
X &  &  & X &  & X & X &  & 37.1\% & 56.1\% & 60.2\%    \\
 & X &  &  &  &  &  &  & 37.1\% & 58.7\% & 73.1\%       \\
X & X &  & X &  & X & X &  & 37.0\% & 60.2\% & 70.7\%   \\
 & X & X &  &  &  &  &  & 36.7\% & 55.9\% & 73.5\%      \\
X & X & X & X &  & X &  &  & 36.6\% & 62.6\% & 68.5\%   \\
 & X &  & X &  &  &  &  & 36.4\% & 58.5\% & 73.7\%      \\
X & X &  & X & X &  & X & X & 36.2\% & 55.4\% & 58.5\%  \\
 & X &  &  &  & X &  &  & 36.0\% & 58.5\% & 68.1\%      \\
X & X &  &  & X & X & X &  & 36.0\% & 61.1\% & 69.4\%   \\
 &  &  &  &  & X &  &  & 36.0\% & 59.8\% & 63.3\%       \\
X & X & X &  &  &  & X &  & 35.9\% & 60.7\% & 70.3\%    \\
 & X & X & X &  &  &  &  & 35.6\% & 55.9\% & 72.6\%     \\
 &  & X &  &  &  & X &  & 35.5\% & 56.6\% & 74.4\%      \\
 & X & X & X &  &  & X &  & 35.4\% & 62.8\% & 73.3\%    \\
X &  & X & X &  &  & X &  & 35.3\% & 56.5\% & 69.2\%    \\
 &  & X & X &  & X &  & X & 34.8\% & 62.2\% & 62.3\%    \\
X & X &  & X &  & X &  &  & 34.6\% & 55.6\% & 61.5\%    \\
 &  & X & X &  &  & X &  & 34.5\% & 57.2\% & 71.5\%     \\
X & X & X & X &  &  & X &  & 34.3\% & 61.2\% & 67.5\%   \\
X & X & X & X & X &  & X &  & 34.2\% & 58.6\% & 64.2\%  \\
X & X & X & X &  &  &  &  & 34.2\% & 55.3\% & 65.3\%    \\
X &  &  & X &  &  & X &  & 33.6\% & 55.3\% & 73.1\%     \\
X &  &  &  &  &  & X &  & 33.3\% & 55.2\% & 72.1\%      \\
X & X & X &  &  &  &  &  & 33.2\% & 56.5\% & 64.4\%     \\
 &  & X &  &  & X &  &  & 32.6\% & 60.6\% & 65.3\%      \\
X & X &  & X &  &  & X &  & 32.5\% & 54.9\% & 67.9\%    \\
 &  & X & X &  & X &  &  & 32.3\% & 62.9\% & 64.4\%     \\
X & X &  &  &  &  & X &  & 32.3\% & 50.9\% & 63.4\%     \\
X & X &  & X &  &  &  &  & 31.6\% & 54.8\% & 62.5\%     \\
X & X & X &  & X & X & X &  & 31.3\% & 57.0\% & 56.8\%  \\
X & X &  &  &  &  &  &  & 30.6\% & 46.9\% & 60.2\%      \\
X &  & X &  &  &  & X &  & 30.1\% & 43.1\% & 66.3\%     \\
 &  & X &  &  &  &  &  & 26.6\% & 29.5\% & 64.8\%       \\
X &  & X & X &  &  &  & X & 26.1\% & 41.7\% & 60.8\%    \\
X &  & X &  &  &  &  &  & 25.7\% & 32.7\% & 63.2\%      \\
X &  & X & X &  &  &  &  & 25.0\% & 40.1\% & 58.8\%     \\
 &  & X & X &  &  &  & X & 24.7\% & 24.7\% & 63.5\%     \\
 &  & X & X &  &  &  &  & 24.7\% & 24.7\% & 63.5\%      \\
X &  &  & X &  &  &  &  & 24.6\% & 24.6\% & 63.4\%      \\
 &  &  & X &  &  &  & X & 24.6\% & 24.6\% & 63.4\%      \\
 &  &  & X &  &  &  &  & 24.6\% & 24.6\% & 63.4\%       \\
X &  &  & X &  &  &  & X & 24.6\% & 24.6\% & 63.4\%     \\

\bottomrule                                             
\end{tabular}%

\end{table}

\begin{table}[ht]
\setlength{\tabcolsep}{3.5pt}
\centering
\scriptsize
\caption{Top 251 - 255 runs of feature combination study.}\label{tab:feature_combination_study_large_results_251_to_255}
\begin{tabular}{ccccccccccc}
\toprule
\multicolumn{8}{c}{Features}& \multicolumn{3}{c}{Classification}  \\
BT & PCE & HR & GL & AT & HU & RT & WS & 7-point Acc & 3-point Acc & 2-point Acc \\
\midrule

 &  &  &  &  &  &  & X & 24.6\% & 27.8\% & 63.1\%       \\
 &  & X &  &  &  &  & X & 24.5\% & 24.5\% & 63.2\%      \\
X &  & X &  &  &  &  & X & 24.5\% & 24.5\% & 63.2\%     \\
X &  &  &  &  &  &  & X & 24.5\% & 24.5\% & 63.1\%      \\
X &  &  &  &  &  &  &  & 24.5\% & 24.5\% & 63.1\%       \\

\bottomrule                                             
\end{tabular}%

\end{table}

\vfill\clearpage

\section{Confusion Matrices}
\label{sec:confusion_matrices}

\begin{figure}[ht]
    \centering
        \includegraphics[width=0.5\textwidth]{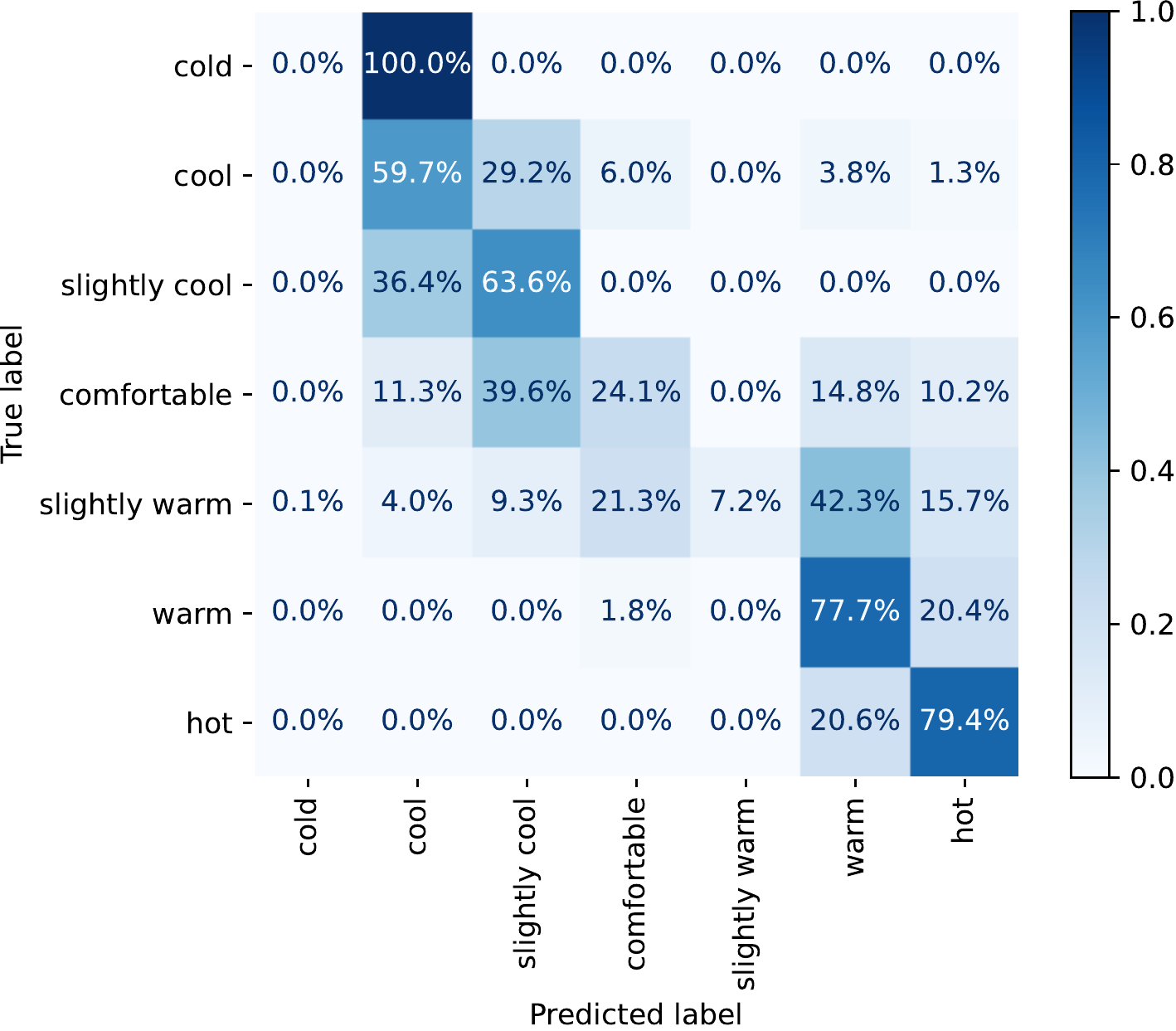}
        \caption{\label{fig:conf_rf_7} Random Forest confusion matrix. \textit{Cold} classes are misclassified entirely, and \textit{Comfortable} classes are rarely classified correctly. Warmer labels are predicted with higher accuracy.}
        \Description{This Figure shows the confusion matrix for the Random Forest classifier. The highest values are mostly located along the matrix diagonal except for the labels 0, -2, and -3, which are more spread around the matrix diagonal. This indicates that cold, cool, and comfortable states were mostly misclassified.}
\end{figure}

\begin{figure}[ht]
    \centering
        \includegraphics[width=0.5\textwidth]{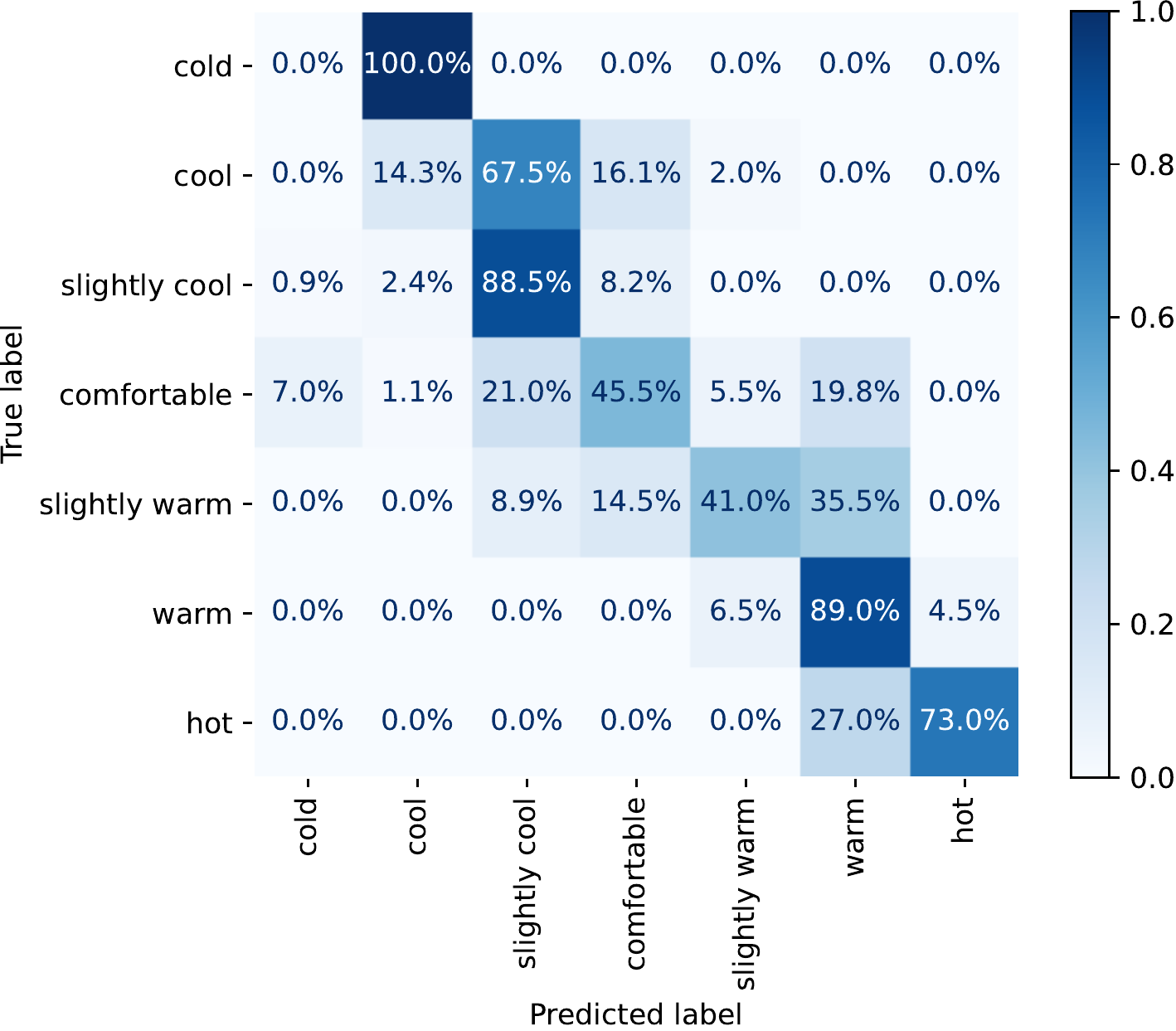}
        \caption{\label{fig:conf_lstm_7} LSTM confusion matrix. \textit{Comfortable} and \textit{Cold} states are mostly wrongly classified. Warmer labels were classified more reliably.}
           \Description{This Figure shows the confusion matrix for the LSTM classifier. The highest values are mostly located along the matrix diagonal except for the labels 0 and -3, which are more spread around the matrix diagonal. This indicates that cold and comfortable states were often misclassified.}
\end{figure}

\begin{figure}[ht]
    \centering
        \includegraphics[width=0.5\textwidth]{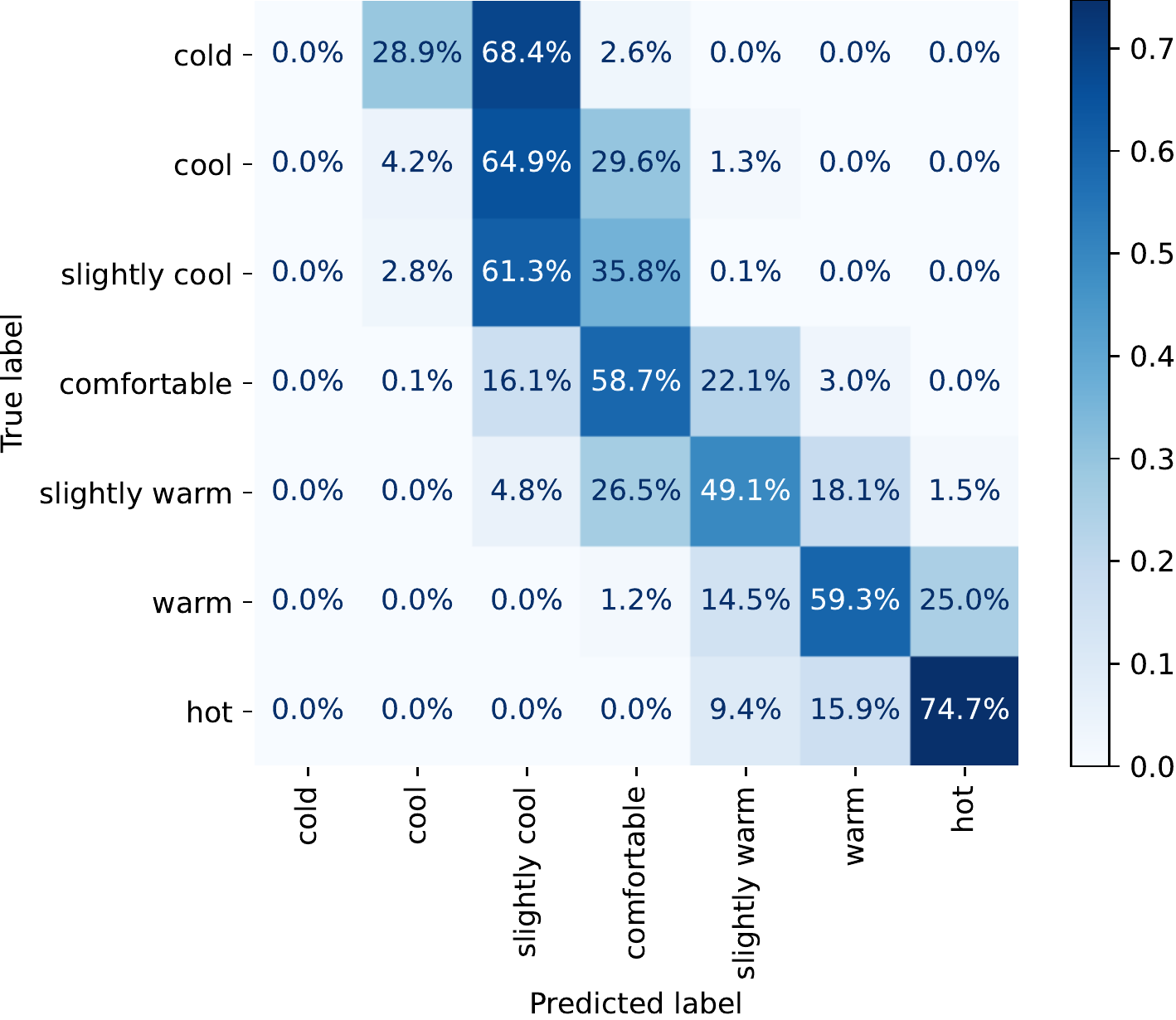}
        \caption{\label{fig:conf_rcnn_7} The CNN-LSTM confusion matrix. Most classes apart from \textit{Hot} states are wrongly classified.}
        \Description{This Figure shows the confusion matrix for the CNN-LSTM classifier. The highest values are mostly spread along the matrix diagonal. This indicates that all states were mostly wrongly classified.}
\end{figure}


\begin{figure*}[ht] 
  \begin{subfigure}{0.3\textwidth}
    \centering
    \includegraphics[width=.8\linewidth]{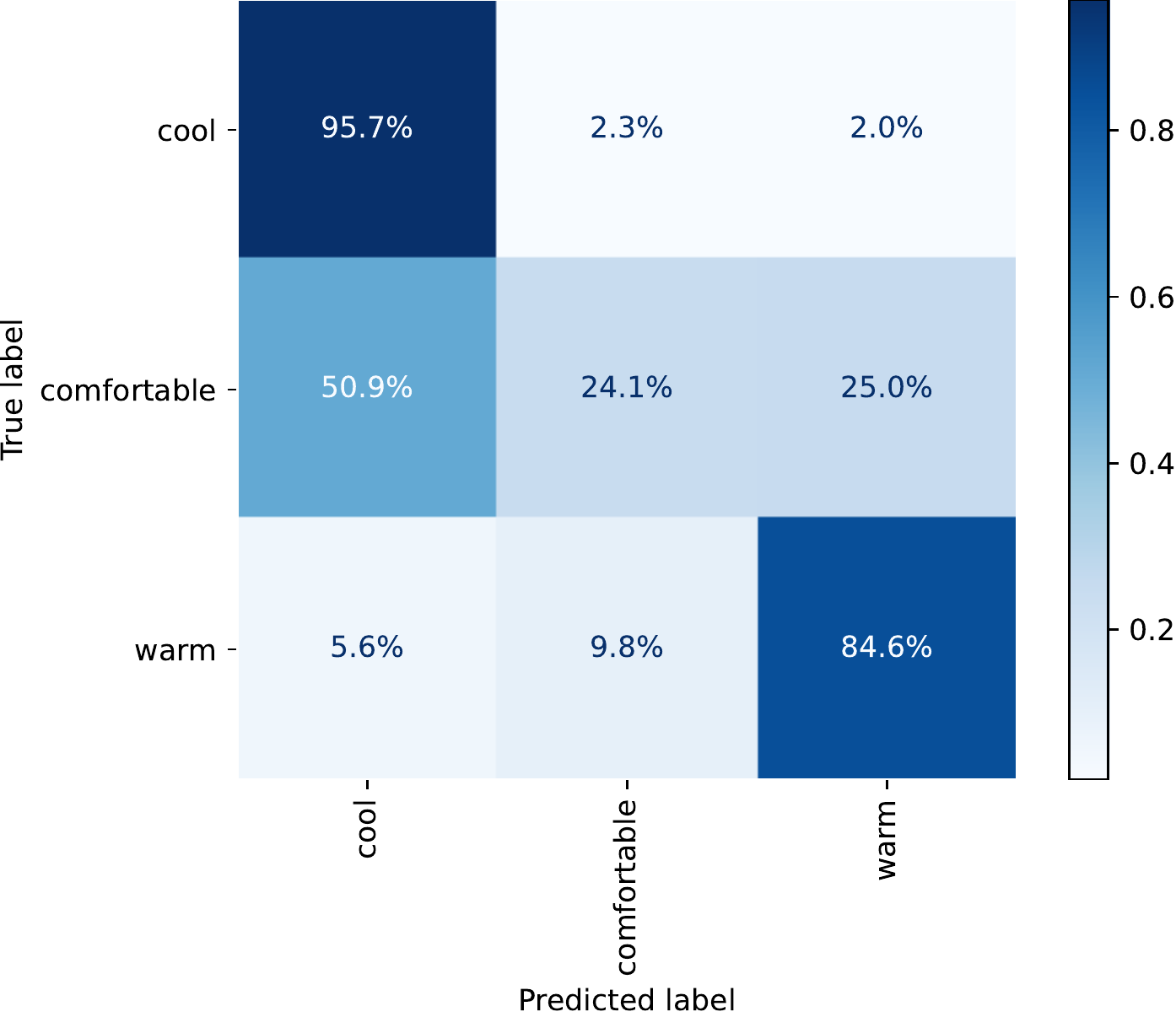}
    \label{fig:1}
  \end{subfigure}%
  \begin{subfigure}{0.3\textwidth}
    \centering
    \includegraphics[width=.8\linewidth]{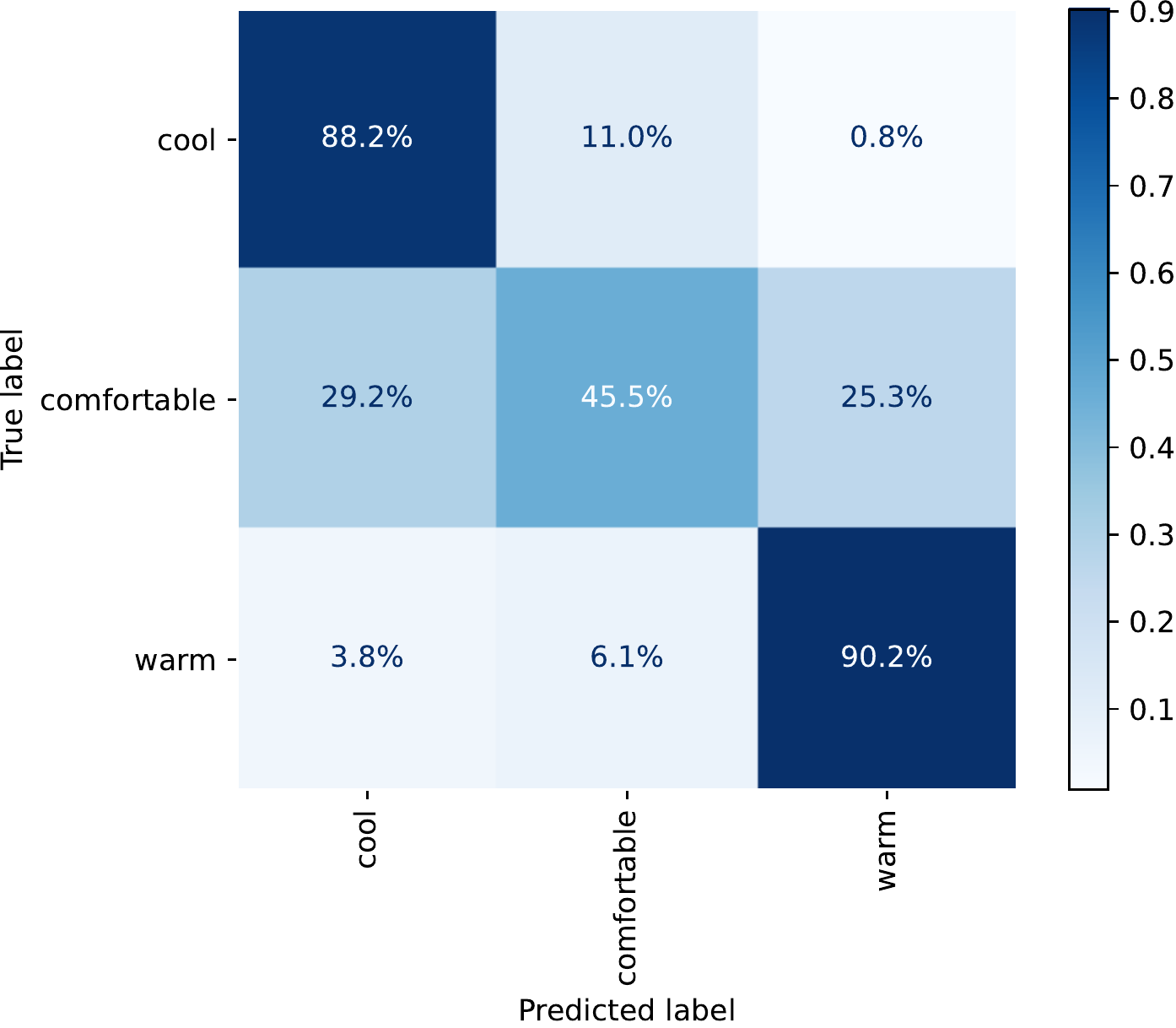}
    \label{fig:2}
  \end{subfigure}
  \begin{subfigure}{0.3\textwidth}\quad
    \centering
    \includegraphics[width=.8\linewidth]{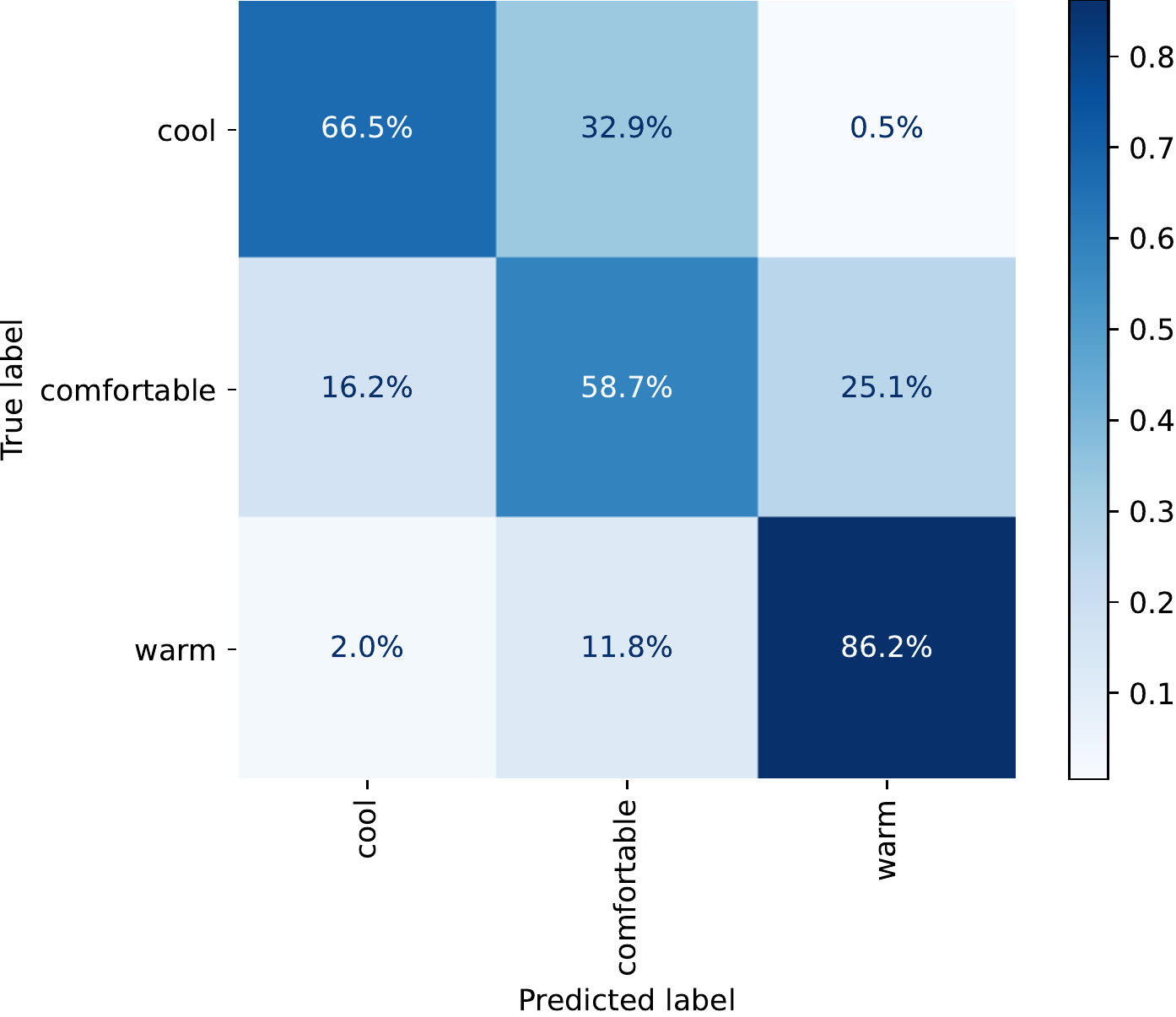}
    \label{fig:3}
  \end{subfigure}
  \medskip

  \begin{subfigure}{0.3\textwidth}
    \centering
    \includegraphics[width=.8\linewidth]{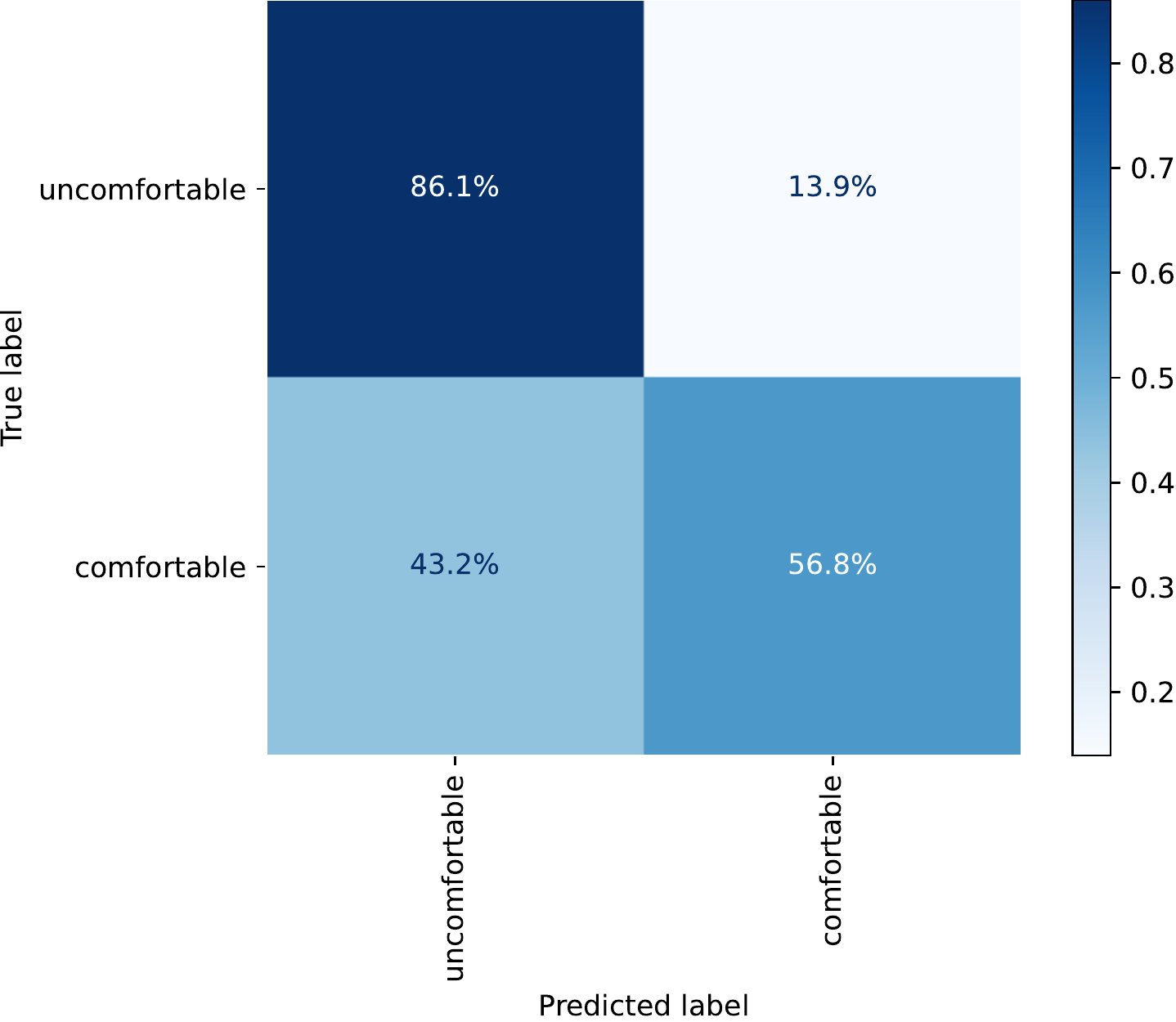}
    \caption{Random Forest}
    \label{fig:4}
  \end{subfigure}
  \begin{subfigure}{0.3\textwidth}
    \centering
    \includegraphics[width=.8\linewidth]{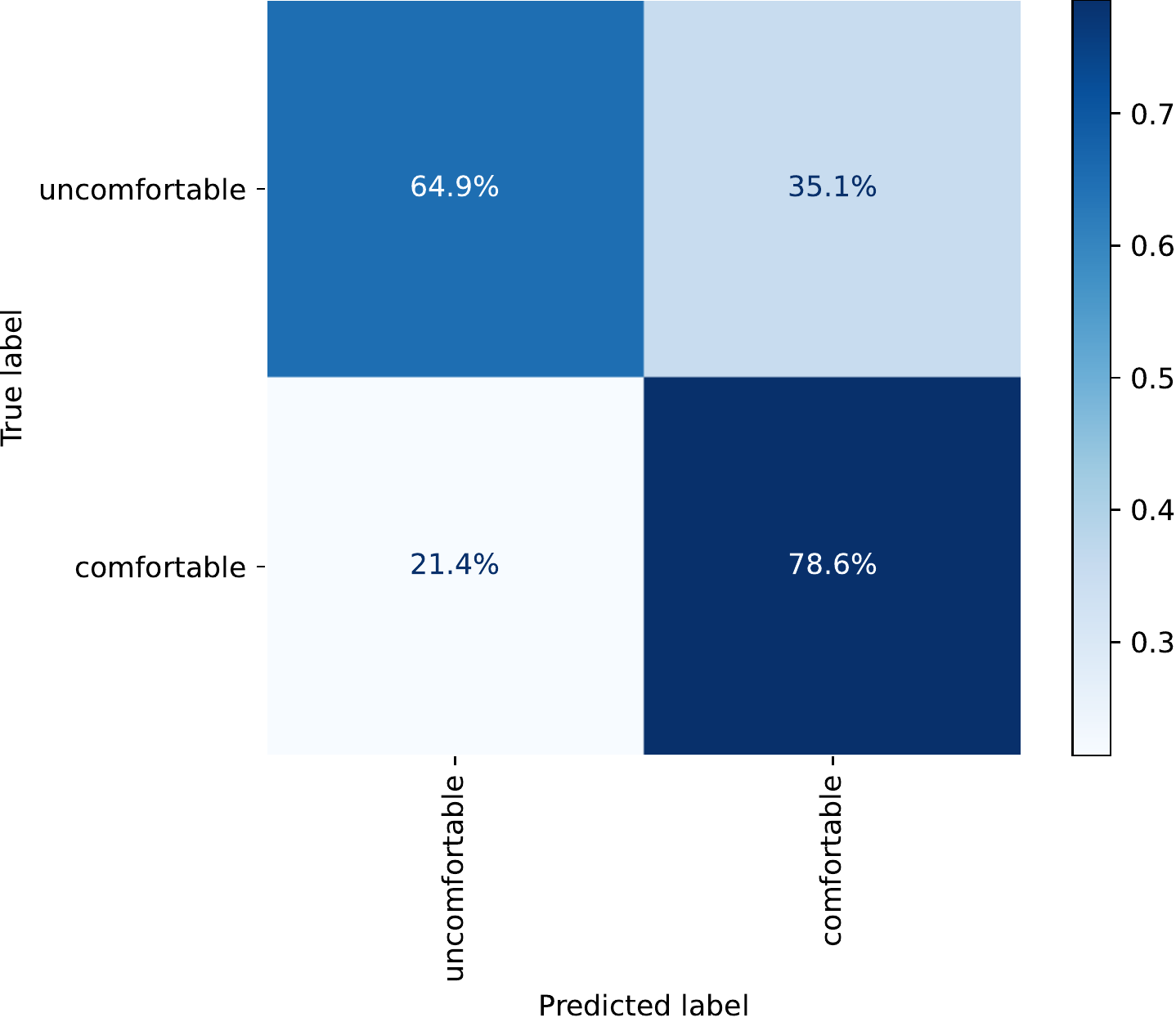}
    \caption{Recurrent Network Classifier}
    \label{fig:5}
  \end{subfigure}
  \begin{subfigure}{0.3\textwidth}
    \centering
    \includegraphics[width=.8\linewidth]{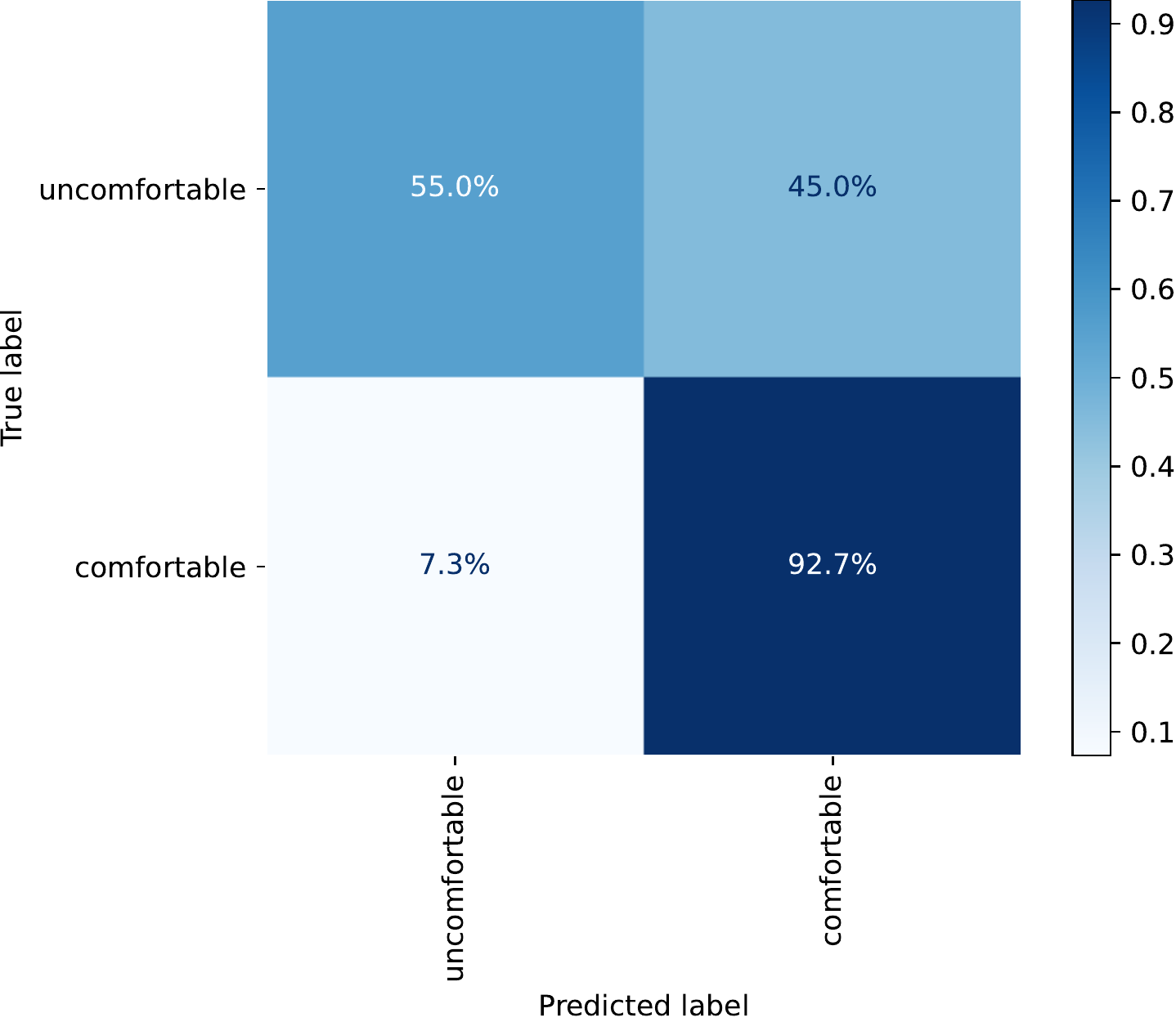}
    \caption{CNN + RNN Classifier}
    \label{fig:6}
  \end{subfigure}
  \caption{Confusion matrices for three-point and two-point classification.}
  \label{fig:images}
  
  \Description{This Figure shows the confusion matrices for the three-point and two-point classification runs grouped by the employed classifier (Random Forest, RNN classifier, and CNN-RNN classifier). The first column on the left includes the confusion matrices for the Random Forest classifier. The 3-class confusion matrix shows that the classes cool and warm were predicted more reliably (cool with 95.7\% and warm with 84.6\%). The comfortable class was only classified correctly 24.1\% of the time. In the 2-class confusion matrix, the uncomfortable class was classified correctly 86.1\% of the time, while the comfortable class was classified correctly only 56.8\% of the time.
  
  The second column in the middle includes the confusion matrices for the RNN classifier. The 3-class confusion matrix shows that the classes cool and warm were predicted reliably (cool with 88.2\% and warm with 90.2\%). The comfortable class was classified correctly 45.5\% of the time. In the 2-class confusion matrix, the uncomfortable class was classified correctly 64.9\% of the time, while the comfortable class was classified correctly only 78.6\% of the time.
  
  The third column on the right includes the confusion matrices for the CNN-RNN classifier. The 3-class confusion matrix shows that the class warm was predicted more reliably (86.2\%) while cool and comfortable classes were predicted less reliably (cool 66.5\% and comfortable 58.7\%). In the 2-class confusion matrix, the uncomfortable class was correctly classified 55.0\% of the time, while the comfortable class was classified correctly 92.7\% of the time.
  }
\end{figure*}

\begin{figure*}[ht]
        \includegraphics[width=0.5\linewidth]{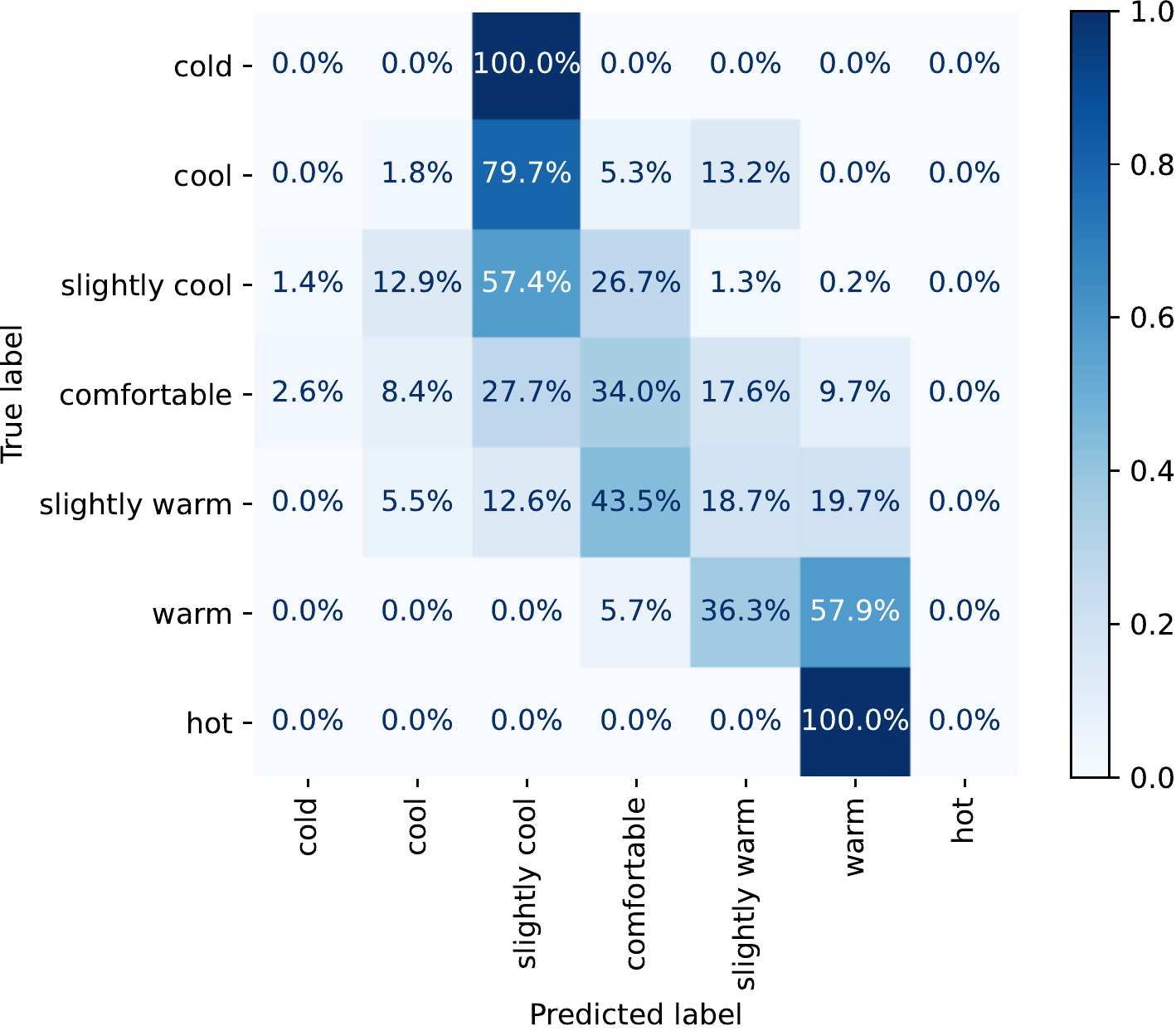}
        \caption{\label{fig:pmv_conf} The confusion matrix resulting from PMV index calculated labels. It can be seen that all classes were mostly wrongly classified using the PMV index. }
           \Description{This Figure shows the confusion matrix for PMV index estimated labels. The values are more broadly spread around the matrix diagonal. This indicates that all states were mostly wrongly classified.}
\end{figure*}
\vfill\clearpage
\section{Thermal Comfort Study Introductory Slides}\label{sec:introduction}
We provide the slides used to explain the procedure and the data logging interface to participants. Most slides included screenshots from the actual data logger GUI to ensure that explanations given during the introduction were supported by visualizations of the actual logging application.


\section*{Supplementary material}

# Thermal Comfort Study

**Introductory Slides**

## Navigation

Once the logging application was started, use **only the marked keys** on the provided **numeric pad to your right.**

Follow the supervisor's instructions and feel free to ask questions if anything is unclear.

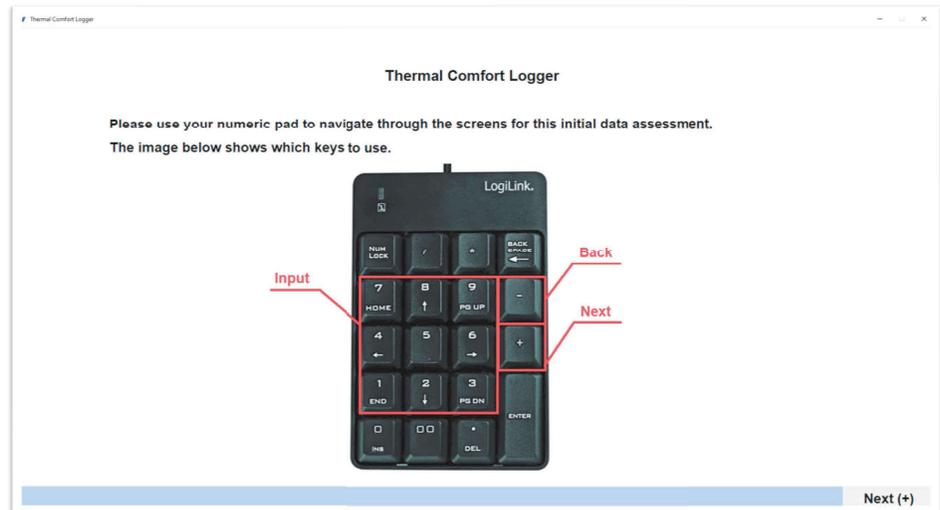

**Thermal Comfort Logger**

*Enter your age:

Questions with an * are
**mandatory** and cannot be skipped.

Back (-)                                                                 Next (+)

---

**Thermal Comfort Logger**

**Which gender do you identify as?

- Male (1)

- Female (2)

- Non-Binary (3)

- Prefer not to tell (4)

Back (-)                                                                 Next (+)

## Thermal Comfort Logger

Enter your weight in kg (e.g. 93,5):

Optional questions can be skipped by
pressing **+** on the numeric pad

Back (-)

Next (+)

## Thermal Comfort Logger

Enter your height in cm (e.g. 153,5):

Back (-)

Next (+)

# Thermal Comfort Logger

Enter your bodyfat percentage (e.g. 15,5):

Back (-)      Next (+)

# Thermal Comfort Logger

*Enter your body temperature (e.g. 36,5):

Back (-)      Next (+)

**Thermal Comfort Logger**

*Did you perform any physical activities/sports within the last hour?

○ No (0)                                                    ○ Yes (1)

Back (-)                                                    Next (+)

---

**Thermal Comfort Logger**

*How much time has passed since your last meal (in full hours):

Back (-)                                                    Next (+)

# Thermal Comfort Logger

**\*Please rate how tired you're feeling at the moment.**

**Not at all**  ○ ○ ○ ○ ○ ○ ○ ○ ○ ○  **Very Tired**
           1    2    3    4    5    6    7    8    9   10

Back (-)     Next (+)

---

# Thermal Comfort Logger

**Enter the ID of the clothing level that describes your current clothing best.**
**Note: Shoes, socks, and underwear are already included in all clothing levels.**

| ID | Description |
|----|-------------|
| 1 | Trousers, short-sleeve shirt |
| 2 | Trousers, long-sleeve shirt |
| 3 | Trousers, long-sleeve sweater |
| 4 | Trousers, long-sleeve vest |
| 5 | Sweatpants, short-sleeve shirt |
| 6 | Sweatpants, long-sleeve shirt |
| 7 | Sweatpants, long sleeve sweater |

**\*Enter your ID here:**

Back (-)     Next (+)

**Thermal Comfort Logger**

*Which item describes your current emotion best?

○ **Anger (1)**

○ **Fear (2)**

○ **Sadness (3)**

○ **Disgust (4)**

○ **Happiness (5)**

○ **Neutral (6)**

○ **Surprise (7)**

○ **Contempt (8)**

| Back (-) | | Next (+) |

---

**Thermal Comfort Logger**

*Which item describes your current state of thermal comfort best?

○ Cold (1)  ○ Cool (2)  ○ Slightly Cool (3)  ○ Comfortable (4)  ○ Slightly Warm (5)  ○ Warm (6)  ○ Hot (7)

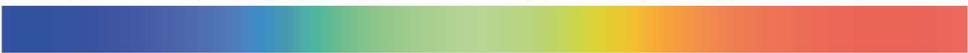

Once you've arrived at this page, do not press submit until the supervisor tells you to.

| Back (-) | | Submit (+) |

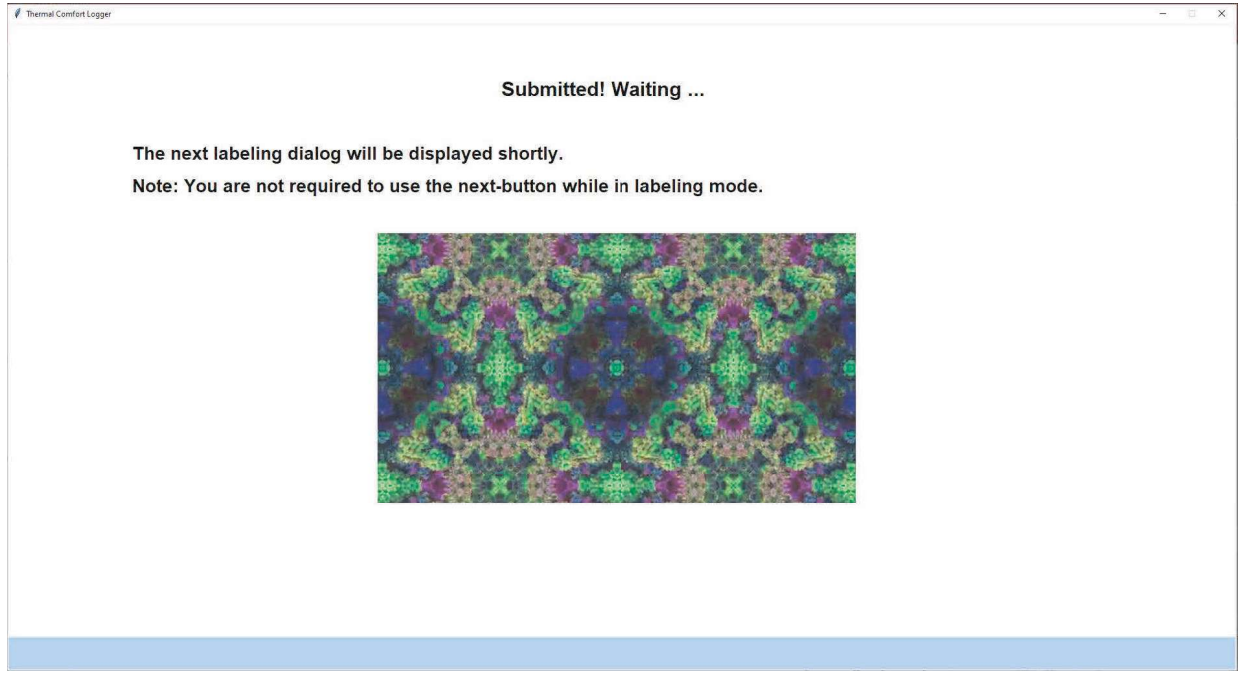

# Labeling

While in labeling mode you'll notice that there are no next/back buttons. **This is intended.**

In this mode, only use the numeric input keys to label, the page will change itself.

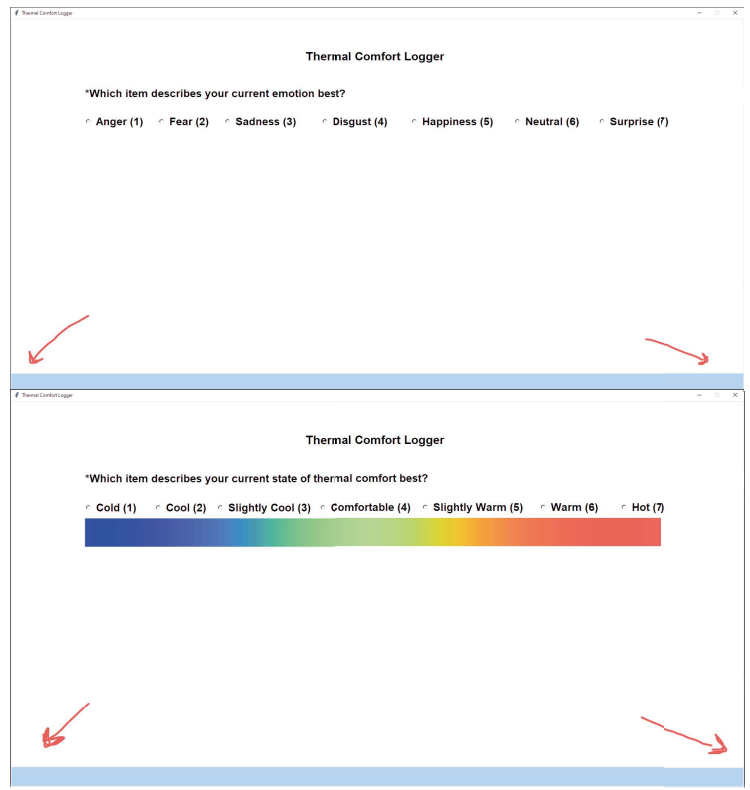

# Other Important Hints

- If you feel unwell during the recording session, let the supervisor know.
- The interface can be displayed in german.
- **Do not use your phone or any other devices.**
- **Do not hold keyboard keys or press repeatedly in case of slow downs.**
- **If your numpad stops working you might've pressed the numlock button. Press it again to activate the numpad.**



\end{document}